\documentclass[draftclsnofoot, onecolumn, romanappendices]{IEEEtran}
\pdfoutput=1

\usepackage{ifpdf}
\usepackage{cite}

\ifCLASSINFOpdf
  \usepackage[pdftex]{graphicx}
  \DeclareGraphicsExtensions{.pdf}
\else
  \usepackage[dvips]{graphicx}
  \DeclareGraphicsExtensions{.eps}

\fi
\usepackage{tikz}
\usetikzlibrary{arrows,shapes,backgrounds,plotmarks}

\usepackage{pgfplots}
\usepackage{pgfplotstable}
\usepackage{booktabs}
\usepackage{colortbl}
\pgfplotsset{grid style={dashed,gray}}

\usepackage{multirow} 

\usepackage[font=footnotesize]{caption}
\usepackage[font=footnotesize]{subcaption}

\usepackage[cmex10]{amsmath}
\usepackage{amssymb}
\usepackage{amsthm}

\usepackage{array}

\ifpdf
\usepackage[pdftex]{hyperref}
    \hypersetup{
    colorlinks=true,%
    citecolor=black,%
    filecolor=black,%
    linkcolor=black,%
    urlcolor=black
  }  
\else

\fi

\tikzstyle{dot}=[circle,draw=gray!80,fill=gray!20,thick,inner
 sep=2pt,minimum size=11pt]
\tikzstyle{vnode} = [circle,draw=gray!80,fill=gray!80,thick,inner
 sep=0pt,minimum size=5pt]
\tikzstyle{cnode} = [draw=gray!80,fill=gray!80,thick,inner
 sep=0pt,minimum size=5pt]

\newtheorem{theorem}{Theorem}

\newtheorem{lemma}{Lemma}

\theoremstyle{definition}

\newtheorem{scheme}{Scheme}
\theoremstyle{remark}

\DeclareMathOperator{\E}{\mathbb{E}}

\DeclareMathOperator{\Bi}{I}
\DeclareMathOperator{\rank}{\text{rk}}
\newcommand{\diff}{\mathit{d}}

\newcommand{\ffield}{\mathbb{F}}

\newcommand{\cmatt}[2]{\zeta^{#1}_{#2}}
\newcommand{\bigO}{\mathcal{O}}

\newcommand{\bG}{\mathbf{G}}
\newcommand{\bB}{\mathbf{B}}
\newcommand{\bX}{\mathbf{X}}
\newcommand{\bY}{\mathbf{Y}}

\newcommand{\bH}{\mathbf{H}}

\newcommand{\degr}{\mathrm{dg}}

\begin{document}
\title{Batched Sparse Codes}
\author{Shenghao~Yang and
  Raymond~W.~Yeung,~\IEEEmembership{Fellow,~IEEE}%
  \thanks{This paper was presented in part at the IEEE International Symposium on Information Theory, Saint Petersburg, Russia, August 2011.}%
  \thanks{This work was supported in part by the National
    Basic Research Program of China Grant 2011CBA00300, 2011CBA00301,
    the National Natural Science Foundation of China Grant 61033001,
    61361136003. This work was partially funded by a grant from the University Grants Committee of the Hong Kong Special Administrative Region (Project No.\ AoE/E-02/08) and Key Laboratory of Network Coding, Shenzhen, China (ZSDY20120619151314964).}%
  \thanks{S. Yang is with the Institute for Theoretical Computer Science, Institute for Interdisciplinary Information Sciences, Tsinghua University, Beijing, China (email: shyang@tsinghua.edu.cn).}%
  \thanks{R. W. Yeung is with the Institute of Network Coding and the Department of Information Engineering, The Chinese University of Hong Kong, N.T., Hong Kong, and with the Key Laboratory of Network Coding Key Technology and Application and Shenzhen Research Institute, The Chinese University of Hong Kong, Shenzhen, China (e-mail: whyeung@ie.cuhk.edu.hk).}%
}

\maketitle

\begin{abstract}

Network coding can significantly improve the transmission rate of communication networks with packet loss compared with routing. However, using network coding usually incurs high computational and storage costs in the network devices and terminals. For example, some network coding schemes require the computational and/or storage capacities of an intermediate network node to increase linearly with the number of packets for transmission, making such schemes difficult to be implemented in a router-like device that has only constant computational and storage capacities.
In this paper, we introduce BATched Sparse code (BATS code), which enables a digital fountain approach to resolve the above issue. BATS code is a coding scheme that consists of an outer code and an inner code.  The outer code is a matrix generation of a fountain code.  It works with the inner code that comprises random linear coding at the intermediate network nodes.  BATS codes preserve such desirable properties of fountain codes as ratelessness and low encoding/decoding complexity. The computational and storage capacities of the intermediate network nodes required for applying BATS codes are independent of the number of packets for transmission. Almost capacity-achieving BATS code schemes are devised for unicast networks, two-way relay networks, tree
networks, a class of three-layer networks, and the butterfly network.  For general networks, under different optimization criteria, guaranteed decoding rates for the receiving nodes can be obtained.

\end{abstract}


\section{Introduction}

One fundamental task of communication networks is to distribute a bulk
of digital data, called a \emph{file}, from a source node to a set of
destination nodes.  We consider this file distribution problem, called
\emph{multicast}, in \emph{packet networks}, in which data packets
transmitted on the network links can be lost due to channel noise,
congestion, faulty network hardware, and so on.

Existing network protocols, for example TCP, mostly use retransmission
to guarantee reliable transmission of individual packets.
Retransmission relies on feedback and is not scalable for multicast
transmission. On the other hand, fountain codes, including LT codes
\cite{lubyLT}, Raptor codes \cite{shokRaptor} and online codes
\cite{maymounkov02}, provide a good solution without relying on
feedback for routing networks, where the intermediate nodes apply
store-and-forward. When using fountain codes, the source node keeps
transmitting coded packets generated by a fountain code encoder and a
destination node can decode the original file after receiving $n$
coded packets, where $n$ typically is only slightly larger than the
number of the input packets, regardless of which $n$ packets are
received.  Fountain codes have the advantages of ratelessness,
universality, and low encoding/decoding complexity.  Taking Raptor
codes as an example, both the encoding and decoding of a packet has
constant complexity.

Routing, however, is not an optimal operation at the intermediate
nodes for multicast. For a general network, the maximum multicast rate
can be achieved only by {\em network coding} \cite{flow}. Network
coding allows an intermediate node to generate and transmit new
packets using the packets it has received.  Linear network
coding was proved to be sufficient for multicast
communications~\cite{linear,alg} and can be realized distributedly by random linear
network coding~\cite{random,chou03,jaggi03,Sanders03}.

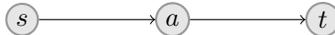
\begin{figure}
  \centering
  \begin{tikzpicture}
     \node[dot] (s) at(-2,0) {$s$};
     \node[dot] (a) at(0,0) {$a$} edge[<-] (s);
     \node[dot] (t) at(2,0) {$t$} edge[<-] (a);
  \end{tikzpicture}
  \caption{Three-node network. Node $s$ is the source node, node $t$
    is the destination node, and node
    $a$ is the intermediate node that does not demand the file. Both
    links are capable of transmitting one packet per use and have a
    packet loss rate $0.2$.}
  \label{fig:three}
\end{figure}

Moreover, routing is not optimal in the presence of packet loss from
the throughput point of view, \emph{even for unicast}. For example,
the routing capacity of the network in Fig.~\ref{fig:three} is $0.64$
packet per use.\footnote{Here one use of a network means the use of
  all network links at most once.} If we allow decoding and encoding
operations at the intermediate node and treat the network as a
concatenation of two erasure channels, we can achieve the rate $0.8$
packet per use by using erasure codes on both links.

The following network coding method has been proved to achieve the
multicast capacity for networks with packet loss in a wide range of
scenarios~\cite{Lun2008,Wu06Tre,Dana2006}. The source node transmits random
linear combinations of the input packets and an intermediate node
transmits random linear combinations of the packets it has received.
Note that no erasure codes are required for each link though packet
loss is allowed. Network coding itself plays the role of end-to-end
erasure codes. A destination node can decode the input packets when it
receives enough coded packets with linearly independent coding
vectors. This scheme is referred to as the \emph{baseline random
  linear network coding scheme (baseline RLNC scheme)}.

The baseline RLNC scheme has been implemented for small number of
input packets, e.g., 32 \cite{Chachulski2007}, but the scheme is
difficult to be implement efficiently when the number of input packets
is large due to the computational and storage complexities and the
coding vector overhead.  Consider transmitting $K$ packets where each
packet consists of $T$ symbols in a finite field.  The encoding of a
packet at the source node takes $\bigO(TK)$ finite field operations,
where $T$ is given and $K$ goes to infinity. 
A finite field operation refers to the addition or multiplication of
two field elements.
An intermediate node
needs to buffer all the packets it has received for network coding, so
in the worst case, the storage cost is $K$ packets and the computation
cost of encoding a packet is $\bigO(TK)$ finite field operations.
Decoding using Gaussian elimination costs on average $\bigO(K^2+TK)$
finite field operations per packet. Though these complexities are
polynomials in $K$, the baseline RLNC scheme is still difficult to
implement for large $K$.

Coding vectors are used in random linear network coding to recover the
linear transformation induced by network coding \cite{random}.  For
transmitting $K$ input packets, the baseline RLNC scheme requires each
packet includes a coding vector of $K$ symbols. Hence, the coding
vector overhead is $K$ symbols per packet or $K/T$ percent. Network
communication systems usually have a maximum value for $T$, e.g.,
several thousands of symbols. Therefore, for large values of $K$, the
coding vector overhead is significant.

In this paper, we study file transmission through networks with packet
loss using network coding.
We hope to build network coding enabled devices with
limited storage and computational capabilities. 
Accordingly, it is desirable for a network coding scheme to
have i) low encoding complexity in the source node and low decoding
complexity in the destination nodes, ii) constant computational
complexity for encoding a packet at an intermediate node and constant
buffer requirement in an intermediate node,\footnote{A constant buffer
  requirement is desirable because one may not know ahead of time the
  size of the file to be transmitted.} iii) small protocol and
coding vector overhead, and iv) high transmission rate.

\subsection{Related Works}
\label{sec:previous}

Existing works mostly use one
of the following two approaches to reduce the computation/storage
complexity and the coding vector overhead. These approaches apply in
the presence or absence of packet loss. The first approach is to use
chunks, each of which is a subset of the input packets. A large file can
be separated into a number of small chunks, and network coding is
applied to each chunk \cite{chou03}. The use of small chunks can
effectively reduces the computational complexity and coding vector
overhead. Therefore this idea is used in many implementations of random
linear network coding in both wireline 
networks~\cite{gkan05,mwang07} and wireless networks \cite{Katti2006,
  Chachulski2007}.  However, the use of chunks introduces the scheduling
issue of chunks since all or a large fraction of the chunks are
required to be decoded individually.  Specifically, sequential scheduling of chunks
requires feedback and is not scalable for multicast, while random
scheduling of chunks requires the intermediate nodes to cache all the
chunks \cite{maym06}. A detailed discussion of the scheduling issues
of chunks can be found in~\cite{yang11ISABEL}.

Further, random scheduling of chunks becomes less efficient when a
fraction of chunks have been decoded.  To resolve this issue of the
random scheduling of chunks, both precoding \cite{maym06} and chunks
with overlapping (chunked codes)
\cite{Silva2009,Heidarzadeh2010,yaoli11} have been considered.
Precoding allows the input packets to be recovered when only a
fraction of all the chunks have been successfully decoded. Chunked
codes use the already decoded chunks to help the decoding of the other
chunks.  A design of chunked codes with precoding using expander graph
is proposed by Tang, et al. \cite{bin12expander}. A technique similar
to chunked codes is Gamma codes \cite{Mahdaviani12,Mahdaviani13},
which can be regarded as chunked codes with parity check constraints
between chunks.  Both the chunked codes mentioned above and the Gamma
codes are fixed rate codes. That is, for a given number of input
packets, only a fixed number of chunks can be generated.

The second approach is to use fountain codes for networks
with coding at the intermediate nodes. The low complexity belief
propagation decoding algorithm of LT/Raptor codes depends on a
suitably chosen
degree distribution.  Since coding at the intermediate nodes changes
the degrees of the coded packets, it is difficult to guarantee that the
degrees of the received packets follow a specific
distribution. Heuristic algorithms have been proposed for special
network topologies (e.g., line networks
\cite{Pakzad2005,gummadi2008relaying}) and special communication
scenarios (e.g., peer-to-peer file sharing \cite{champel10,Thomos11}),
but these solutions are difficult to be extended to general network
settings and require the intermediate nodes to have a buffer size that
increases linearly with the number of packets for transmission.

In addition to the above two approaches, there are techniques focusing
on certain specific issues or scenarios. For example, an error
correction code based approach is proposed by Jafari et al. to reduce
the coding vector overhead \cite{Jafari09c}. This approach puts a
limit on the number of packets that can be combined together, but does
not put the decoding complexity into consideration.  Link-by-link
feedback can be used to reduce the storage at the intermediate nodes
\cite{Fragouli07f,Keller08b,Sundar08a}. Jaggi et al. have proposed a
binary permutation matrix based approach to reduce the complexity of
the finite field operations in linear network coding \cite{Jaggi06l}.

\subsection{Our Solution}

In this paper, we propose an efficient linear network coding solution
based on a new class of codes called \emph{BATched Sparse (BATS)
  codes}, which extend fountain codes to incorporate random linear
network coding.  A BATS code consists of an inner code and an outer
code over a finite field. The outer code is a matrix generalization of
a fountain code, and hence rateless. The outer code encodes the file
to be transmitted into \emph{batches}, each containing $M$ packets.
When the batch size $M$ is equal to 1, the outer code reduces to a
fountain code.  The inner code applies a linear transformation on
each batch and is represented by the linear transfer matrices of the
batches.  The inner code is formed by \emph{linear network coding}
performed at the intermediate network nodes with the constraint that
only packets belonging to the same batch can be combined inside the
network. 
The property of the inner code preserves the degrees of the batches so
that an efficient belief propagation (BP) decoding algorithm can
be used to jointly decode the outer code and the inner code.

BATS codes are suitable for any network that allows linear network
coding at the intermediate nodes. BATS codes are robust against
dynamical network topology and packet loss since the end-to-end
operation remains linear. Moreover, BATS codes can operate with small
finite fields. In contrast, most existing random linear network coding
schemes require a large field size to guarantee a full rank for the
transfer matrix.  For BATS codes, the transfer matrices of the batches
are allowed to have arbitrary rank deficiency. We demonstrate the
applications of BATS codes in unicast networks, two-way relay
networks, tree networks, the butterfly network and peer-to-peer file
distribution.

BATS codes resolve the feedback issue of sequential scheduling of the
chunk-based approach: Feedback is not required for sequential
scheduling of batches due to the rateless property.  BATS codes also
resolve the degree distribution issue of the fountain-code-based
approach since the inner code of BATS codes (induced by linear network
coding) does not change the degrees of the batches.  When applying
BATS codes, the encoding of a packet by the outer code costs
$\bigO(TM)$ finite field operations, where $T$ and $M$ are given and
$K$ goes to infinity. An intermediate node uses $\bigO(TM)$ finite
field operations to recode a packet, and an intermediate node is
required to buffer only $\bigO(M)$ packets for tree networks,
including the three-node network in Fig.~\ref{fig:three}.  BP decoding
of BATS codes costs on average $\bigO(M^2+TM)$ finite field operations
per packet. The coding vector overhead of a BATS code is $M$ symbols
per packet. Note that all these requirements for BATS codes are
constant for $K$, the total number of packets for transmission.

The (empirical) rank distribution of the transfer matrices of the
batches plays an important role in BATS codes. The optimization of the
outer code depends only on the rank distribution.  We use density
evolution to analyze the BP decoding process of BATS codes, and obtain
a sufficient and a necessary condition for BP decoding recovering
a given fraction of the input packets with high probability. For given
rank distributions, a degree distribution for a BATS code can be
obtained by solving an optimization problem induced by the sufficient
condition.

For any inner code with rank distribution $(h_0,h_1,\ldots,h_M)$, we
verify theoretically for certain cases and demonstrate numerically for
general cases that the outer code with BP decoding achieves rates
very close to the expected rank $\sum_i i h_i$, the theoretical upper
bound on the achievable rate of the code in packets per batch. For
unicast erasure networks, BATS codes with BP decoding can achieve the
min-cut capacity asymptotically when both $M$ and $T$ tend to
infinity.  This can be extended to multicast erasure networks when all
the destination nodes have the same empirical rank distribution (which
is rare in practice).

When the destination nodes have different empirical rank
distributions, we can optimize the degree distribution for various
criteria, and obtain a set of guaranteed rate tuples for BP decoding.
However, there is no guarantee in general that with this degree
distribution, the rate of BP decoding at each destination node can
achieve the expected rank for that node. For a given batch size, we
can obtain numerically the percentage of the expected rank that is
achievable for all possible rank distributions by using one degree
distribution. For example, the percentage is at least 52.74 for batch
size 16. When the possible empirical rank distributions are in a
smaller set, a better degree distribution achieving higher rates can
be found.

BP decoding is an efficient but not the only way to decode BATS
codes. Gaussian elimination can be used to continue the decoding when
BP decoding stops. A better algorithm, especially for relative small
$K$, e.g., several ten thousands, is inactivation decoding, which
has been used for Raptor codes \cite{inactivation}.  Other techniques
like finite-length analysis and precodes fine-tuned for inactivation
decoding are also required to design efficient BATS codes with finite
block lengths. We discuss these techniques briefly in this paper with
reference to existing literature for details.

\subsection{Organization of this Paper}

BATS codes are formally introduced in Section~\ref{sec:smc}. The
belief propagation decoding of BATS codes is analyzed in
Section~\ref{sec:ana}. A necessary and a sufficient condition such
that the BP decoding stops with a given fraction of the input packets
recovered is obtained in Theorem~\ref{the:1}, which is proved in
Section~\ref{sec:prove}.  The degree distribution optimizations and
the achievable rates of BATS codes are discussed in
Section~\ref{sec:deg}.  The degree distribution optimizations of BATS
codes for multiple rank distributions is discussed in
Section~\ref{sec:opt2}. The necessary techniques for the design of the
outer codes and decoding algorithms with good finite length
performance are discussed in Section~\ref{sec:finite}.  Examples of
how to use BATS codes in networks, as well as the design of the inner
code of a BATS code, are given in
Section~\ref{sec:example}. Concluding remarks are in
Section~\ref{sec:conc}.

\section{BATS Codes}
\label{sec:smc}

In this section, we discuss the encoding and decoding of BATS codes.
Consider encoding $K$ input packets, each of which has $T$ symbols in
a finite field $\ffield$ with size $q$.  A packet is denoted by a
column vector in $\ffield^T$.  The rank of a matrix $\mathbf{A}$ is
denoted by $\rank(\mathbf{A})$.  In the following discussion, we
equate a set of packets to a matrix formed by juxtaposing the packets
in this set.  For example, we denote the set of the input packets by
the matrix
\begin{equation*}
  \mathbf B = \begin{bmatrix}b_1,b_2,\cdots,b_K\end{bmatrix},
\end{equation*}
where $b_i$ is the $i$th input packet.  On the other hand, we also
regard $\mathbf{B}$ as a set of packets, and so, with an abuse of
notation, we also write $b_i\in\mathbf B$, $\bB'\subset \bB$, etc.

\subsection{Encoding of Batches}
\label{sec:genbatch}

Let us first describe the outer code of a BATS code, which generates
code packets in batch. (We also call the outer code itself the BATS
code when the meaning is clear from the context.)  A \emph{batch} is a
set of $M$ coded packets generated from a subset of the $K$ input
packets.  For $i=1,2,\ldots$, the $i$th batch $\bX_i$ is generated
from a subset $\bB_i\subset\bB$ of the input packets by the operation
\begin{equation*}
  \bX_i = \bB_i \bG_i, %
\end{equation*}
where $\bG_i$, a matrix with $M$ columns, is called the
\emph{generator matrix} of the $i$th batch. We call the packets in
$\bB_i$ the contributors of the $i$th batch.  The formation of $\bB_i$
is specified by a \emph{degree distribution}
$\Psi=(\Psi_0,\Psi_1,\cdots,\Psi_K)$ as follows: 1) sample the distribution
$\Psi$ which returns a \emph{degree} $d_i$ with probability $\Psi_{d_i}$;
2) uniformly at random choose $d_i$ input packets to form $\bB_i$.
The design of $\Psi$ is crucial for the performance of BATS code,
which will be discussed in details in this paper.

The generator matrix $\bG_i$ has dimension $d_i\times M$ and can be
generated randomly. Specifically, $\bG_i$ is the instance of a
$d_i\times M$ random matrix $G_i$, in which all the components are
independently and uniformly chosen at random.  Such a random matrix is
also called a \emph{totally random matrix}.  We analyze BATS codes
with random generator matrices in this paper.  Random generator
matrices do not only facilitate analysis but are also readily
implementable. For example, $\bG_i$, $i=1,2,\cdots$ can be generated
by a pseudorandom number generator and can be recovered at the
destination nodes by the same pseudorandom number generator.

The generator matrices can also be designed deterministically. For
example, when $d_i\leq M$, we can pick $\bG_i$ such that $\rank(\bG_i)
= d_i$. When $d_i > M$, we can use the generator matrix of an MDS code
as the generator matrix of the $i$th batch. But we would not analyze
the performance of such transfer matrices in this paper.

When $M=1$, the above batch encoding process becomes the encoding of LT
codes. We are interested in this paper the case $M>1$. There are no
limits on the number of batches that can be generated. So BATS code
can be used as a rateless code.

\begin{figure}
  \centering
  \begin{tikzpicture}[scale=0.7]
     \node[vnode, label=above:$b_1$] (v1) at(-4,2) {};
     \node[vnode, label=above:$b_2$] (v2) at(-2,2) {};
     \node[vnode, label=above:$b_3$] (v3) at(0,2) {};
     \node[vnode, label=above:$b_4$] (v4) at(2,2) {};
     \node[vnode, label=above:$b_5$] (v5) at(4,2) {};
     \node[vnode, label=above:$b_6$] (v6) at(6,2) {};

     \node[cnode, label=left:$\bG_1$] (c1) at(-4,0) {} 
     edge [<-] (v1) edge [<-] (v3); 
     \node[cnode, label=left:$\bG_2$] (c2) at(-2,0) {} edge [<-]
     (v4) edge [<-] (v5); 
     \node[cnode, label=left:$\bG_3$] (c3) at(0,0) {} edge
     [<-] (v6) edge [<-] (v2); 
     \node[cnode, label=left:$\bG_4$] (c4) at(2,0) {}
     edge [<-] (v3) edge [<-] (v4) edge [<-] (v5); 
     \node[cnode, label=left:$\bG_5$] (c5)
     at(4,0) {} edge [<-] (v2) edge [<-] (v1) edge [<-] (v4);

     \node[cnode] (d1) at(-4,-2) {} edge [<-] node[swap,auto] {$\bH_1$} (c1);
     \node[cnode] (d2) at(-2,-2) {} edge [<-] node[swap,auto] {$\bH_2$} (c2);
     \node[cnode] (d3) at(0,-2) {} edge [<-] node[swap,auto] {$\bH_3$} (c3);
     \node[cnode] (d4) at(2,-2) {} edge [<-] node[swap,auto] {$\bH_4$} (c4);
     \node[cnode] (d5) at(4,-2) {} edge [<-] node[swap,auto] {$\bH_5$} (c5);
  \end{tikzpicture}
  \caption{Tanner graph for the inner and the outer code of a BATS
    code. Nodes in the first row are the variable nodes representing
    the input packets. Nodes in the second row are the check nodes
    representing the batches generated by the outer code. Nodes in the
    third row are the check nodes representing the batches processed
    by the inner code.}
  \label{fig:enc}
\end{figure}
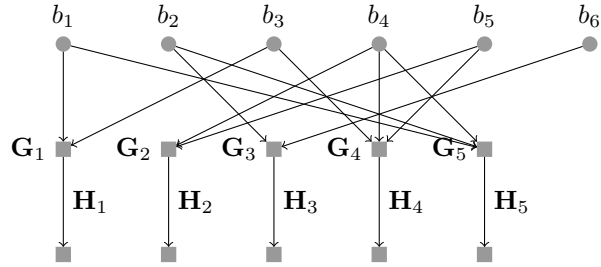

The batch encoding process can be described by a Tanner graph.  The
Tanner graph has $K$ \emph{variable nodes}, where variable node $i$
corresponds to the $i$th input packet $b_i$, and $n$ \emph{check
  nodes}, where check node $j$ corresponds to the $j$th batch $\bX_j$.
Check node $j$ is connected to variable node $i$ if $b_i$ is a
contributor of $\bX_j$. Associated with each check node $j$ is the
generator matrix $\bG_j$.  Fig.~\ref{fig:enc} illustrates an example
of a Tanner graph for encoding batches.

\subsection{Transmission of Batches}
\label{sec:trans}

Now we turn to the inner code of a BATS code. The batches generated by
the outer code are transmitted from a network employing network coding
to multiple destination nodes. We assume that the end-to-end
transformation of each batch is a linear operation. Fix a
destination node. Let $\bH_i$ be the transfer matrix of the $i$th
batch and $\bY_i$ be the output (received) packets of the $i$th
batch. We have
\begin{equation}\label{eq:trans}
  \bY_i = \bX_i \bH_i = \bB_i\bG_i\bH_i.
\end{equation}
The number of rows of $\bH_i$ is $M$. The number of columns of $\bH_i$
corresponds to the number of packets received for the $i$th batch, which may
vary for different batches and is finite.  We assume that $\bH_i$ is
known for decoding. In linear network coding, this knowledge can be
obtained at the destination nodes through the coding vectors in the
packet headers.

In other words, we assume that a received packet of a destination node
cannot be the linear combinations of the packets of more than one
batch from the same BATS code. To obtain such received packets, we may
assume that an intermediate node can only apply network coding on
packets of the same batch.\footnote{It is possible that network coding
  between packets of different batches is applied locally so that the
  coded packets of different batches in an intermediate node can be
  decoded directly at the nodes in the next hop.}  Packet loss and
dynamical network topology are allowed during the network
transmission.  The benefits of applying network coding within batches
includes
\begin{itemize}
\item The network coding complexity at an intermediate node is
  $\bigO(MT)$ finite field operations per packet, which does not
  depend on $K$.
\item The coding vector overhead is bounded by $M$. When
  the packet length $T$ is sufficiently larger than $M$, this overhead is
  negligible.
\end{itemize}
Moreover, since packets from different batches will not be
encoded together, it is not necessary to keep all the batches in an
intermediate node for the purpose of network coding.

We call the the network coding scheme at the intermediate network
nodes the \emph{inner code} of a BATS code.  The transfer matrices of batches
are determined jointly by the inner code and the network
topology between the source node and the destination node.  Under the
principle that only packets of the same batch can be recoded, we have
a lot of freedom in designing the inner code, including how to
manage the buffer content, how to schedule the transmission of
batches/packets, and how to use the feedback messages. The design of
the inner code is closely related to the network
topology. We will use several typical network topologies to
demonstrate how to design the inner code such that the
benefit of BATS codes is maximized (see Section~\ref{sec:example}).

The empirical rank distribution of the transfer matrices is an
important parameter for the design of BATS codes. The empirical rank
distribution determines the maximum achievable rate of the outer code
and provides sufficient information to design nearly optimal outer
codes.  Since many network operations are random, e.g., random linear
network coding, random packet loss pattern and network topology
dynamics, the transfer matrices are also random matrices. Consider
$\bH_i$ as the instance of a random matrix $H_i$.  The operation of
the network on the batches in \eqref{eq:trans} can be modeled as a
channel with input $X_i$ and output $Y_i = X_i H_i$, $i=1,2,\ldots$,
where the instance of $H_i$, regarded as the state of the channel, is
known by the receiver.  This channel model is called a \emph{linear
  operator channel (LOC) with receiver side channel state
  information}. Similar channel models has been studied without the
channel state information \cite{koetter08j,silva08c}.  Unless
otherwise specified, receiver side channel state information is
assumed for all the LOCs discussed in this paper.  The LOC is not
necessary to be memoryless since $H_i$, $i=1,2,\ldots$ are not assumed
to be independent. With receiver side channel state information, the
capacity of the LOC can be easily characterized. Consider that
\begin{equation*}
  \lim_{n\rightarrow \infty} \frac{\sum_{i=1}^n \rank(H_i)}{n} \stackrel{P}{\longrightarrow} \bar h.
\end{equation*}
We can check that channel capacity of the above channel is upper
bounded by $\bar h$ and the upper bound can be achieved by random
linear codes \cite{yang10bf}.  As a channel code for the LOC, the
maximum achievable rate of the outer code of a BATS code is bounded by
$\bar h$ for any inner code with average rank of the transfer matrices
converging to $\bar h$.  From the above analysis, we should design the
inner code to maximize $\bar h$.  Define the
\emph{design coding rate} of a BATS code as $K/n$.  As we will show in
Section~\ref{sec:deg}, for a given empirical rank distribution $(h_0,
h_1,\ldots,h_M)$, we have an outer code that can achieve a rate very
close to $\sum_i ih_i$.

\subsection{Belief Propagation Decoding}
\label{sec:dec}

\begin{figure}
  \centering
  
  \begin{tikzpicture}[scale=0.7]
     \node[vnode] (v1) at(-4,2) {};
     \node[vnode] (v2) at(-2,2) {};
     \node[vnode, label=above:$b_k$] (v3) at(0,2) {};
     \node[vnode] (v4) at(2,2) {};
     \node[vnode] (v5) at(4,2) {};
     \node[vnode] (v6) at(6,2) {};

     \node[cnode] (c1) at(-4,0) {} edge [<-] (v1) edge [<-] (v3);
     \node[cnode] (c2) at(-2,0) {} edge [<-] (v4) edge [<-] (v5);
     \node[cnode, label=below:$\bG_{i}\bH_{i}$] (c3) at(0,0) {} edge [<-] (v6) edge [<-] (v3) edge [<-] (v2);
     \node[cnode, label=below:$\bG_{j}\bH_{j}$] (c4) at(2,0) {} edge [<-] (v3) edge [<-] (v4) edge [<-] (v5);
     \node[cnode] (c5) at(4,0) {} edge [<-] (v2) edge [<-] (v1) edge [<-] (v4);
  \end{tikzpicture}
  \caption{A decoding graph. Nodes in the first row are the variable nodes
    representing the input packets. Nodes in the second row are the
    check nodes representing the batches.}
  \label{fig:dec}
\end{figure}
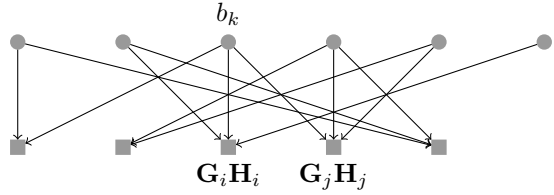

A destination node tries to decode the input packets using $\bY_i$ and the
knowledge of $\bG_i$ and $\bH_i$ for $i=1,2,\ldots,n$. The decoding is
equivalent to solving the system of linear equations formed by
\eqref{eq:trans} for $i=1,\ldots,n$. Solving the system of linear
equations using Gaussian elimination has high computational cost when
$K$ is large. We propose a belief propagation (BP) based low complexity
decoding algorithm for BATS codes. The BP decoding process
is better described using the bipartite graph in Fig.~\ref{fig:dec},
which is the same as the encoding graph in Fig.~\ref{fig:enc} except
that the two stages of the encoding are combined together and
the overall transfer matrix $\bG_i\bH_i$ is associated with each check node $i$.

A check node $i$ is called decodable if $\rank(\bG_i\mathbf{H}_i)$ is
equal to the degree of the $i$th batch $d_i$. If check node $i$ is
decodable, then $\bB_{i}$ can be
recovered by solving the linear system of equations $\bY_{i} = \bB_{i}
\bG_{i} \bH_{i}$, which has a unique solution since $\rank(\bG_{i}
\bH_{i}) = d_{i}$.  After decoding the ${i}$th batch, we recover the
$d_{i}$ input packets in $\bB_{i}$. Next, we substitute the values of
these input packets in $\bB_{i}$ in the undecoded batches.  Consider that $b_k$ is
in $\bB_{i}$. If variable node $k$ has only one edge that connects
with check node~$i$, just remove variable node~$k$. If variable node
$k$ also connects check node $j\neq {i}$, then we further reduce the
degree of check node $j$ by one and remove the row in $\bG_{j}$ corresponding to
variable node $k$.  In the decoding graph, this is equivalent to first
removing check node~$i$ and its neighboring variable nodes, and then
for each removed variable node update its neighboring check nodes.  We
repeat this decoding-substitution procedure on the new graph until no
more check nodes are decodable.

One of the main tasks of this paper is to understand the performance of BATS
code under BP decoding, which will be discussed in
Section~\ref{sec:ana}-\ref{sec:deg}. 

\subsection{Computational Complexity}
\label{sec:complexity}

In the following computational complexity, the unit is a
finite field operation. Suppose that $T$ and $M$ are given, and $K$
and $n$ are
the variables that tend to infinite in the big O notation.

To generate a batch of degree $d$, we combine $d$ packets together $M$
times. So generating a batch with degree $d$ costs $\bigO(TMd)$ finite
field operations. Thus the
encoding complexity of $n$ batches is $\bigO(TM\sum_{i=1}^n d_i)$, 
which converges to $\bigO(TMn\bar{\Psi})$  finite
field operations when $n$ is large, where
$\bar{\Psi}=\sum_d d\Psi_d$ is the average degree.

Let $k_i = \rank(\bH_i)$ and let $k_i'$ be the rank of $\bG_i\bH_i$
when check node $i$ is decodable.  It is clear that $k_i'\leq k_i\leq
M$. By the definition of the decodability of a check node, $k_i'$ is
also the degree of check node $i$ when it is decodable.  Since the
degree of a check node tends to decrease at each step of the decoding
process, we have $k_i' \leq d_i$.  The decoding processing involves
two parts: the first part is the decoding of the decodable check
nodes, which costs $\bigO(\sum_i k_i'^3 + T\sum_ik_i'^2)$
finite field operations; the second part is the updating of the
decoding graph, which costs $\bigO(T \sum_i(d_i-k_i')M)$ finite field
operations.  So the total complexity is $\bigO(\sum_i k_i'^3 +
T\sum_ik_i'^2 + T \sum_i(d_i-k_i')M)$, which can be simplified to
$\bigO(n M^3 + TM\sum_id_i)$.  When $n$ is large, the complexity
converges to $\bigO(M^3n + TMn\bar{\Psi})$  finite
field operations. Usually, $T$ and
$\bar{\Psi}$ is larger than $M$ and the second term is dominant.

\subsection{Precoding}
\label{sec:precod}

From the above complexity analysis, we see that the expected degree $\bar{\Psi}$
affects the encoding/decoding complexity.  Consider that we want to
recover all the $K$ input symbols using $n$ batches with 
probability at least $1-1/K^c$ for some positive constant $c$. Similar to
the analysis of LT codes (cf. \cite[Proposition 1]{shokRaptor}), no
matter what decoding algorithm is applied the expected
degree is lower bounded by $c'\frac{K}{n}\log(K)$ for some positive
constant $c'$.

\begin{figure}
  \centering
  \begin{tikzpicture}[scale=0.7]
    \node[vnode] (b0) at (2,4) {};
    \node[vnode] (b1) at (0,4) {};
    \node[vnode] (b2) at (-2,4) {};
    \node[vnode] (b3) at (-4,4) {};

     \node[cnode] (v1) at(-4,2) {} edge[<-] (b3);
     \node[cnode] (v2) at(-2,2) {} edge[<-] (b2);
     \node[cnode] (v3) at(0,2) {} edge[<-] (b1);
     \node[cnode] (v4) at(2,2) {} edge[<-] (b0);
     \node[cnode] (v5) at(4,2) {} edge[<-] (b0) edge[<-] (b1) edge[<-] (b2);
     \node[cnode] (v6) at(6,2) {} edge[<-] (b1) edge[<-] (b2) edge[<-] (b3);

     \node[cnode, label=left:$\bG_1$] (c1) at(-4,0) {} edge [<-] (v1) edge [<-] (v3);
     \node[cnode, label=left:$\bG_2$] (c2) at(-2,0) {} edge [<-] (v4) edge [<-] (v5);
     \node[cnode, label=left:$\bG_3$] (c3) at(0,0) {} edge [<-] (v6) edge [<-] (v2);
     \node[cnode, label=left:$\bG_4$] (c4) at(2,0) {} edge [<-] (v3) edge [<-] (v4) edge [<-] (v5);
     \node[cnode, label=right:$\bG_5$] (c5) at(4,0) {} edge [<-] (v2) edge [<-] (v1) edge [<-] (v4);

     \node[cnode] (d1) at(-4,-2) {} edge [<-] node[swap,auto] {$\bH_1$} (c1);
     \node[cnode] (d2) at(-2,-2) {} edge [<-] node[swap,auto] {$\bH_2$} (c2);
     \node[cnode] (d3) at(0,-2) {} edge [<-] node[swap,auto] {$\bH_3$} (c3);
     \node[cnode] (d4) at(2,-2) {} edge [<-] node[swap,auto] {$\bH_4$} (c4);
     \node[cnode] (d5) at(4,-2) {} edge [<-] node[swap,auto] {$\bH_5$} (c5);

  \end{tikzpicture}
  \caption{Precoding of BATS codes. Nodes in the first row
    represent the input packets. Nodes in the second row represent the
    intermediate packets generated by the precode. }
  \label{fig:scheme}
\end{figure}
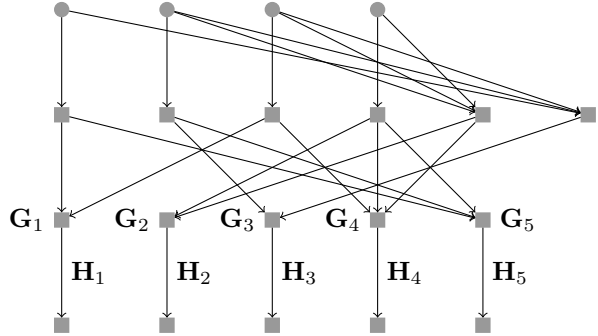

To reduce the $\log(K)$ term in the bound of the expected degree, the
precoding technique of Raptor codes can be applied.  That is, before
applying the batch encoding process in Section~\ref{sec:genbatch}, the
input packets are first encoded using a traditional erasure code
(called a \emph{precode}).  The batch encoding process is applied on
the intermediate input packets generated by the precode. If the belief
propagation decoding of the BATS code can recover a given fraction of
the intermediate input packets, the precode is capable of recovering
the original input packets in face of a fixed fraction of erasures.
Fig. \ref{fig:scheme} illustrates a BATS code with a systematic
precode.

Though BATS codes can be used without precode, we will only study the
design of BATS codes with precode in this paper. Redefine $K$ as the number of
intermediate input packets after precoding. If the precode is designed
to recover the original input packets from at least $(1-\eta)K$
intermediate input packets, the number of original input packets
should be less than or equal to $(1-\eta)K$.  We will see from
Section~\ref{sec:deg} that if we want to recover $(1-\eta)$ of the
(intermediate) input packets by the BP decoding, we only need a degree
distribution with the maximum degree $D < M/\eta$.  When the
normalized design rate $(1-\eta)\frac{K}{nM}$ converges to a constant
value, we see that the encoding and BP decoding complexity are
$\bigO(TKM)$ and $\bigO(KM^2 + TKM)$, respectively.

\bigskip

In the following four sections, we will discuss the analysis and
design of the outer code for given (empirical) rank distributions of
the transfer matrices. Readers can skip this part and go directly to
Section~\ref{sec:example} to see examples of BATS codes applications.

\section{Analysis of BP Decoding}
\label{sec:ana}

Some existing methods for analyzing the BP decoding of erasure codes
can be modified to analyze the BP decoding of BATS codes. In this
paper, we adopt the differential equation approach by Wormald
\cite{wormald99} that has been used by Luby et al. \cite{luby01} to
analyze Tornado codes (see also
\cite{RU08} for an analysis of LDPC codes over erasure channel). %

Compared with the analysis of fountain codes, BATS codes have a
relatively complex decoding criteria that involves both the degree and
the rank value of a check node. In addition to the evolution of the
degrees of the check nodes, the evolution of the ranks of the check
nodes also needs to be tracked in the decoding analysis.

\subsection{Random Decoding Graph}
\label{sec:rd}

Consider a BATS code with $K$ input symbols and $n$ batches. 
Fix a degree distribution
$\Psi=(\Psi_0,\Psi_1,\cdots,\Psi_D)$, where $D$ is the maximum integer
such that $\Psi_D$ is nonzero.  Assume that $D=\bigO(M)$. The
feasibility of this assumption will be justified later. 
The decoder observes a random graph as well as the associated
generator and transfer matrices. The probability model of these
objects is implied in the encoding of BATS codes
described in the last section. Here we explicitly describe this model
for the sake of the analysis.

Denote by $\degr_i, i=1,\ldots, n$ a sequence of i.i.d random
variables each of which follows the distribution $\Psi$. 
Denote by $\mathcal{T}$ a Tanner graph with $K$ variable nodes and
$n$ check nodes. The $i$th check node of $\mathcal{T}$ has degree
$\degr_i$. For a given degree $d$, a check node connects to $d$
variable nodes chosen uniformly at random. Therefore the probability
$\Pr\{\mathcal{T}=T | \{\degr_i=d_i\} \}$ can be fixed. 
The generator matrix $G_i$ of check node $i$ is a
$\degr_i\times M$ totally random matrix, i.e., its components are uniformly
i.i.d. Conditioning on a 
sequence of degree, the Tanner graph and
the generator matrices of the BATS code are obtained independently. 

Let $H_i$ be the transfer matrix associated with check node $i$.  Note
that $H_i, i=1,\ldots,n$ may not be independent. We do not need to
make any assumption on the distribution of $H_i, i=1,\ldots,n$, except that
the empirical distributions of the transfer matrix ranks converge in
probability to a probability vector $h = (h_0,\ldots, h_M)$.
Specifically, for $k=0,\ldots, M$ let
\begin{equation*}
  \pi_k \triangleq \frac{|\{i: \rank(H_i)=k\}|}{n}.
\end{equation*}
Note that $\pi_k$ depends on $n$.
We assume that the convergence of the matrix ranks satisfies
\begin{equation}\label{eq:osp3}
|\pi_k - h_k| =  \bigO(n^{-1/6}), \quad 0\leq k \leq M,
\end{equation}
with probability at least $1 - \gamma(n)$, where $\gamma(n) = o(1)$, i.e., 
there exists a constant $c$ such that for all sufficiently large $n$,
\begin{equation*}
  \Pr\{|\pi_k - h_k|  < cn^{-1/6}, \ 0\leq k \leq M\} > 1-\gamma(n),
\end{equation*}
and
\begin{equation*}
  \lim_{n\rightarrow \infty} \gamma(n) = 0.
\end{equation*}
As an example of valid transfer matrices, 
$\{H_i\}$ are i.i.d. and $\rank(H_i)$ follows the
distribution $h$.  
Hereafter, we call the probability vector $h$ the rank distribution
(of the transfer matrix).
We assume that the transfer matrices are
independent of the generation of batches.

Denote by $\text{BATS}(K, n, \Psi, h)$ the random vector $(\{\degr_i,
G_i, H_i\}_{i=1}^n,
\mathcal{T})$. 
The decoder observes an instance of $\text{BATS}(K, n, \Psi, h)$ with probability
\begin{IEEEeqnarray*}{rCl}
  \IEEEeqnarraymulticol{3}{l}{\Pr\big\{\{\degr_i=d_i,
G_i=\mathbf{G}_i, H_i=\mathbf{H}_i\}_{i=1}^n, \mathcal{T} = T \big\} } \quad \\
  & = & 
  \left(\prod_i \Psi_{d_i}\right) \Pr\big\{\mathcal{T} = T |
    \degr_i=d_i, i=1,\ldots,n \big \}
  \left(\prod_{i} \Pr\{G_i=\mathbf{G}_i| \degr_i=d_i\}\right)
  \Pr\big\{H_i=\mathbf{H}_i, i=1,\ldots,n \big\}.
\end{IEEEeqnarray*}
The decoding runs on an instance of $\text{BATS}(K, n, \Psi,
h)$ and we will look at the convergence of the decoding performance.

We will analyze the decoding performance of $\text{BATS}(K,
n, \Psi, h)$ with a random decoding strategy. In each decoding step,
an edge $(U,V)$ with degree equal to the rank is uniformly chosen,
where $U$ is a check node and $V$ is a variable node. Since check node
$U$ has degree equal to the rank, variable node $V$ is
decodable. Variable node $V$, as well as all the edges connected to
it, are removed in the decoding graph. For each check node connected
to variable node $V$, three operations are applied: 1) the degree is
reduced by $1$; 2) the row in the generator matrix corresponding to
the variable node $V$ is removed; and 3) the rank is updated
accordingly.  The decoding process stops when there is no edge with
degree equal to the rank.  The following decoding analysis is based on
this random decoding strategy. In the decoding process described in
the last section, decoding a check node with degree equal to the rank
can recover several variable nodes simultaneously. Note that for a
given instance of the decoding graph, both strategies will reduce the
decoding graph to the same residual graph when they stop (see the
discussion in Appendix~\ref{sec:layered}).

\subsection{Edge Perspective}

We call $\rank(G_iH_i)$ the
\emph{rank} of check node $i$.
Define the following two regions of the degree-rank pair:
\begin{IEEEeqnarray*}{rCl}
  \bar{\mathcal{F}} & \triangleq & \{(d,r): 1\leq r \leq M, r \leq d \leq D\}, \\
  \mathcal{F} & \triangleq & \{(d,r): 1\leq r \leq M, r < d \leq D\}.
\end{IEEEeqnarray*}
We see that $\bar{\mathcal{F}} = \mathcal{F} \cup \{(r,r),r=1,\ldots,M\}$.
A check node with rank zero does not help the decoding, so we do not
include $(d,0)$ in $\bar{\mathcal{F}}$ and $\mathcal{F}$.  To analyze the
decoding process, we use the degree-rank distribution of the edges
defined as follows.  An edge is said to be of degree $d$ and rank $r$ if it is
connected to a check node with degree $d$ and rank $r$. 
Let $R_{d,r}$ be the number of edges of degree $d$ and rank $r$.
Define the \emph{degree-rank distribution of the edges} as
\begin{equation*}
  \bar R \triangleq (R_{d,r}, (d,r)\in \bar{\mathcal{F}}).
\end{equation*}
Note that $R_{d,r}/d$ gives the number of nodes with degree $d$ and rank $r$.

Using the property of totally random matrix and some counting
techniques in projective space \cite{andrews76,gadouleau10}, we have 
\begin{equation}\label{eq:cp1}
  \Pr\{\rank(G_iH_i)=r| \degr_i=d, \rank(H_i) = k\} = \cmatt{d,k}{r}
  \triangleq \frac{\cmatt{d}{r}\cmatt{k}{r}}{\cmatt{r}{r}q^{(d-r)(k-r)}}
\end{equation}
where
\begin{equation*}
\cmatt{m}{r}\triangleq \left\{\begin{array}{ll}
    (1-q^{-m})(1-q^{-m+1})\cdots(1-q^{-m+r-1}) &
    r>0, \\ 1 & r=0. \end{array} \right.
\end{equation*}
Let
\begin{equation}\label{eq:rdr}
  \rho_{d,r} = d\Psi_d \sum_{k=r}^M \cmatt{d,k}{r} h_k.
\end{equation}
The value $n\rho_{d,r}$ is the expected number of edges of degree $d$ and
rank $r$ in the decoding graph when the rank of a transfer matrix is chosen
according to the probability vector $h$ independently.
The following lemma shows that $R_{d,r}/n$ converges in
probability to $\rho_{d,r}$ as $n$ goes to infinity.

\begin{lemma}\label{lemma:init}
  With probability at least $1- (\gamma(n) +
  2MD\exp(-2n^{2/3}))$,
  \begin{equation*}%
    \left|\frac{{R}_{d,r}}{n} - \rho_{d,r}\right| = \bigO(n^{-1/6}), \quad (d,r)\in \bar{\mathcal{F}}.
  \end{equation*}
\end{lemma}
\begin{IEEEproof}
  Consider the instances of decoding graphs with $\{\pi_k\}$
  satisfying \eqref{eq:osp3}. 
  By the assumption on $\{\pi_k\}$, this will decrease the bound by at
  most $\gamma(n)$.  With an abuse of notation, we treat $\{\pi_k\}$
  as an instance satisfying \eqref{eq:osp3} in the following of this
  proof, i.e.,   the decoding graph has $n\pi_k$ check nodes with
  transfer matrix rank $k$. 
  
  By \eqref{eq:cp1}, the expected number of
  check nodes with degree $d$ and rank $r$ is
  \begin{equation*}
    \sum_{k=r}^M n \pi_k \Psi_d \cmatt{d,k}{r} = n\Psi_d \sum_{k=r}^M\pi_k\cmatt{d,k}{r}.
  \end{equation*}
  Applying Hoeffding's inequality, with probability at least $1- 2MD
  \exp(-2n^{2/3})$,
  \begin{equation}\label{eq:osp1}
   \left|\frac{{R}_{d,r}}{dn} - \Psi_d \sum_{k=r}^M \pi_k \cmatt{d,k}{r} \right| < n^{-1/6} ,\quad (d,r)\in \bar{\mathcal{F}}. 
  \end{equation}
  Then,
  \begin{IEEEeqnarray*}{rCl}
    \left|\frac{{R}_{d,r}}{n} - \rho_{d,r}\right|
    & = & 
    \left|\frac{{R}_{d,r}}{n} - d\Psi_d \sum_{k=r}^M \pi_k \cmatt{d,k}{r} + d\Psi_d \sum_{k=r}^M \pi_k \cmatt{d,k}{r} - d\Psi_d \sum_{k=r}^M h_k \cmatt{d,k}{r}\right| \\
    & \leq  & 
    \left|\frac{{R}_{d,r}}{n} - d\Psi_d \sum_{k=r}^M \pi_k \cmatt{d,k}{r} \right| + d\Psi_d\sum_{k=r}^M |\pi_k-h_k| \cmatt{d,k}{r}.
  \end{IEEEeqnarray*}
  By \eqref{eq:osp1}, under the condition in \eqref{eq:osp3}, we have 
  \begin{equation*}
    \left|\frac{{R}_{d,r}}{n} - \rho_{d,r}\right| = \bigO(n^{-1/6})
  \end{equation*}
  with probability at least $1- 2MD \exp(-2n^{2/3})$. 

  The proof is completed by substracting the probability that
  $\{\pi_k\}$ does not satisfy \eqref{eq:osp3}.
\end{IEEEproof}

\subsection{Density Evolution}
\label{sec:de}

Consider the evolution of $\text{BATS}(K, n, \Psi, h)$ during
the decoding process.  Time $t$ starts at zero and increases by one
for each variable node removed by the decoder. 
During the decoding, some of the random variables we defined in the
previous two subsections will be analyzed as random processes. 
We denote by $\degr_i(t)$ the degree of the $i$th
check node in the residual graph at time $t$, and $G_i(t)$ the
corresponding generator matrix, where $\degr_i(0)=\degr_i$ and $G_i(0)=G_i$.
For $(d,r)\in
\bar{\mathcal{F}}$ let $R_{d,r}(t)$
denote the number of edges in the residual graph of degree $d$ and
rank $r$ at time $t\geq 0$ with $R_{d,r}(0)=R_{d,r}$.  

\begin{figure}
  \centering
  \tikzstyle{dot}=[circle,draw=gray!80,fill=gray!20,thick,inner
 sep=0pt,minimum size=5pt]
  \begin{tikzpicture}[scale=0.95]
    \foreach \k/\d in {1/0, 2/1, 3/2, 4/3, 5/4, 6/5, 7/6, 8/7}
    \foreach \r/\s in {1/0, 2/1, 3/2, 4/3, 5/4}
	{  
          \node[dot] (\k\r) at (\k,-\r) {};
          \ifnum \k>1
            \ifnum \r<\k
              \draw[->] (\k\r) -- (\d\r);
              \ifnum \r>1
                \draw[->] (\k\r) -- (\d\s);
              \fi
            \fi
            \ifnum \r=\k
              \draw[->] (\r\r) -- (\d\d);
            \fi
          \fi
          \ifnum \d<\r
            \breakforeach
          \fi
        }
    \node[above] at (11.north) {$d=1$};
    \foreach \k in {2, 3, 4, 5, 6, 7, 8}
    {
      \node[above] at (\k1.north) {$\k$};
    }
    \node at (9,-1) {$r=1$};
    \foreach \r in {2, 3, 4,5}
    {
      \node at (9.3,-\r) {$\r$};
    }

    \node at (5.4, -2.9) {$\alpha_{6,2}$};
    \node at (5.9, -2.55) {$\bar\alpha_{6,2}$};
    \node at (5.8, -3.5) {$\bar\alpha_{6,3}$};
  \end{tikzpicture}
  \caption{State transition diagram for $M=5$ and $D=8$. Each node in
    the graph represent a degree-rank pair. In each step, 
    if the check node connects to the decoded variable node, its
    state changes according to the direction of the outgoing edges of its
    current state. The label on an edge shows the probability that a
    direction is chosen.}
  \label{fig:dd}
\end{figure}
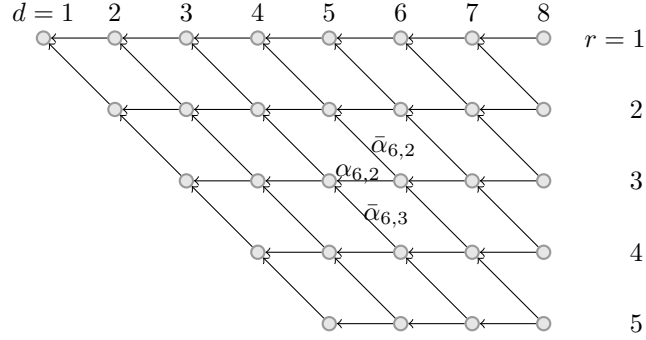

Upon removing a neighboring variable node of a check node with degree $d$ and
rank $r$, the degree of the check node will change to $d-1$.
The rank of the check node may remain unchanged or may change to
$r-1$. Regarding a degree-rank pair as a state, the
state transition of a check node during the decoding process is
illustrated in Fig.~\ref{fig:dd}, where the transition probability is
characterized in the following lemma.

\begin{lemma} \label{lemma:29s0}
  For any check node $i$ and any $ (d,r)\in
  \bar{\mathcal{F}}$,
\begin{IEEEeqnarray*}{rCl}
  \Pr\{\rank(G_i(t+1)H_i) = r \big | \rank(G_i(t)H_i) = r, \degr_i(t+1) = d - 1,
  \degr_i(t) = d\} & = &
  \frac{1-q^{-d+r}}{1-q^{-d}} \triangleq \alpha_{d,r}, \\
  \Pr\{\rank(G_i(t+1)H_i) = r-1 \big | \rank(G_i(t)H_i) = r, \degr_i(t+1) = d - 1,
  \degr_i(t) = d\} & = & 1 - \alpha_{d,r} \triangleq \bar \alpha_{d,r}. 
\end{IEEEeqnarray*}  
\end{lemma}
\begin{IEEEproof}
  We omit the index $i$ in the proof to simplify the notation.
  We have for $k\geq r$,
  \begin{IEEEeqnarray*}{rCl}
    \IEEEeqnarraymulticol{3}{l}{\Pr\{\rank(G(t+1)H) = r \big |
      \rank(G(t)H) = r, \degr(t+1) = d - 1, \degr(t) = d, \rank(H)=k\}} \\
    & = &  \Pr\{\rank(G(t)H) = r | \rank(G(t+1)H) = r, \degr(t+1) = d - 1, \degr(t) = d,
    \rank(H)=k\} \times \\
    & & \times
    \frac{\Pr\{\rank(G(t+1)H) = r | \degr(t+1) = d - 1, \degr(t) = d,
      \rank(H)=k\}}{\Pr\{\rank(G(t)H) = r | \degr(t+1) = d - 1,
      \degr(t) = d, \rank(H)=k\}} \\
    & = & 
    q^{r-k} \frac{\cmatt{d-1,k}{r}}{\cmatt{d,k}{r}} \\
    & = & 
    \frac{1-q^{-d+r}}{1-q^{-d}},
  \end{IEEEeqnarray*}
  where the second equality follows from \eqref{eq:cp1} and the fact
  that $G_i$ is totally random.
  The proof is completed by multiplying $\Pr\{\rank(H)=k|\rank(G(t)H)
  = r, \degr(t+1) = d - 1, \degr(t) = d\}$ on both sides of the above
  equality and taking summation over all $k\geq r$.
\end{IEEEproof}

Assume that the decoding process has not stopped. At time
$t$, we have $K-t$ variable nodes left in the residual graph, and an
edge with degree equal to the rank is uniformly chosen to be
removed. Let
\begin{equation*}
  \bar R(t)\triangleq (R_{d,r}(t):(d,r)\in\bar{\mathcal{F}}).
\end{equation*} 
The random process $\{\bar
R(t)\}$ is a Markov chain, which suggests a straightforward 
approach to compute all the transition probabilities in the Markov
chain. However, this approach leads to
a complicated formula. Instead of taking this approach, we work out
the expected change $R_{d,r}(t+1) - R_{d,r}(t)$ explicitly for all
$t\geq 0$.  Let
\begin{equation*}
  R_{0}(t)=\sum_{r=1}^M R_{r,r}(t).
\end{equation*}
We do not need to study the behavior of $R_{r,r}(t)$
for individual values of $r$ since $R_0(t)$ is sufficient to
determine when the decoding process stops. Specifically, the decoding
process stops as soon as $R_0(t)$ becomes zero.

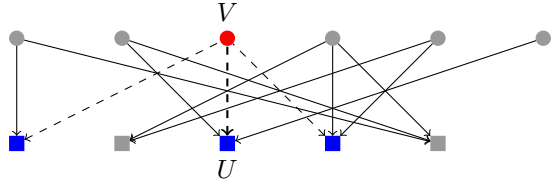
\begin{figure}
  \centering
  
  \begin{tikzpicture}[scale=0.7]
     \node[vnode] (v1) at(-4,2) {};
     \node[vnode] (v2) at(-2,2) {};
     \node[vnode, red, label=above:$V$] (v3) at(0,2) {};
     \node[vnode] (v4) at(2,2) {};
     \node[vnode] (v5) at(4,2) {};
     \node[vnode] (v6) at(6,2) {};

     \node[cnode, blue] (c1) at(-4,0) {} edge [<-] (v1) edge [<-, dashed] (v3);
     \node[cnode] (c2) at(-2,0) {} edge [<-] (v4) edge [<-] (v5);
     \node[cnode, blue, label=below:$U$] (c3) at(0,0) {} edge [<-] (v6) edge
     [<-, thick, dashed] (v3) edge [<-] (v2);
     \node[cnode, blue] (c4) at(2,0) {} edge [<-, dashed] (v3) edge [<-] (v4) edge [<-] (v5);
     \node[cnode] (c5) at(4,0) {} edge [<-] (v2) edge [<-] (v1) edge [<-] (v4);
  \end{tikzpicture}
  \caption{A decoding graph. Edge $(U,V)$ is to be removed at time $t$.}
  \label{fig:densityevolution}
\end{figure}

\begin{lemma}\label{lemma:3}
  For any constant $c\in (0,1)$,
  as long as $t \leq c K$ and $R_0(t)>0$, we have 
\begin{IEEEeqnarray}{rCl}
  \E[R_{d,r}(t+1)-R_{d,r}(t)|\bar R(t)]
  & = & (\alpha_{d+1,r}R_{d+1,r}(t) + \bar \alpha_{d+1,r+1}R_{d+1,r+1}(t)-R_{d,r}(t))%
  \frac{d}{K-t},\ (d,r)\in \mathcal{F}, \label{eq:d9} 
\end{IEEEeqnarray}
and
\begin{IEEEeqnarray}{rCl}
  \E[R_0(t+1) - R_0(t)|\bar R(t)]
  & = & \frac{\sum_r r \alpha_{r+1,r}R_{r+1,r}(t)}{K-t} - \frac{R_0(t)}{K-t} - 1 + \bigO(1/K). \label{eq:d99c}
\end{IEEEeqnarray}
\end{lemma}
\begin{IEEEproof}
  Fix a time $t\geq 0$. %
  With an abuse of
  notation, we treat $\bar R(0),\ldots,\bar R(t)$ as instances in the
  proof, i.e., the values of these random vectors are fixed.
  Let $(U,V)$ be the edge chosen to be removed at
  time $t$, where $V$ is the variable node and $U$
  is the check node, according to the random decoding algorithm described in Section~\ref{sec:rd}.  Note that $V$ is uniformly distributed among all
  variable nodes and $U$ must be a check node with degree equal to the
  rank at time $t$. See the illustration in
  Fig.~\ref{fig:densityevolution}.

  Let $N_{d,r}$ be the number of check nodes which has degree $d$ and
  rank $r$ at time $t$ and has degree $d-1$ at time $t+1$.  Let
  $N_{d,r}^{+}$ (resp. $N_{d,r}^{-}$) be the number of check nodes which has
  degree $d$ and rank $r$ at time $t$ and has degree $d-1$ and rank
  $r$ (resp. $r-1$) at time $t+1$. Clearly, $N_{d,r}^{+}+N_{d,r}^{-} = N_{d,r}$.  
  The difference $R_{d,r}(t+1)-R_{d,r}(t)$ can then be expressed as 
  \begin{equation}
    R_{d,r}(t+1)-R_{d,r}(t)  = d (N_{d+1,r}^{+} + N_{d+1,r+1}^{-} - N_{d,r}). \label{eq:trenddr}
  \end{equation}

  The probability that a check node with degree $d$ and rank
  rank $r$, $d>r$, connects to the variable node $V$ at time $t$ is 
  $d/(K-t)$. Therefore, when $d>r$,
  \begin{equation*}
    N_{d,r} \sim \text{Binom}\left(\frac{R_{d,r}(t)}{d}, \frac{d}{K-t}\right).
  \end{equation*}
  As we characterize in Lemma~\ref{lemma:29s0}, for a check node with degree $d$ and rank
  rank $r$ connecting to the variable node $V$ at time $t$, its degree
  will become $d$ (resp. $d-1$) with probability $\alpha_{d,r}$
  (resp. $\bar \alpha_{d,r}$) at time $t+1$. So when $d>r$, 
  \begin{IEEEeqnarray*}{rCl}
    N_{d,r}^{+} & \sim & \text{Binom}\left(\frac{R_{d,r}(t)}{d},
      \alpha_{d,r}\frac{d}{K-t}\right), \\
    N_{d,r}^{-} & \sim & \text{Binom}\left(\frac{R_{d,r}(t)}{d}, \bar\alpha_{d,r}\frac{d}{K-t}\right).
  \end{IEEEeqnarray*}
The expectation in \eqref{eq:d9} is obtained by taking expectation on
\eqref{eq:trenddr}. 

To verify
\eqref{eq:d99c}, note that $N_{r,r}^{+}=0$ and hence
$N_{r,r}^{-}=N_{r,r}$. Then we have
\begin{IEEEeqnarray*}{rCl}
  R_0(t+1) - R_0(t) & = & \sum_{r}\left( R_{r,r}(t+1) - R_{r,r}(t)\right) \\
  & = & \sum_{r} r N_{r+1,r}^{+} - \sum_r N_{r,r}. \IEEEyesnumber \label{eq:trenddr0}
\end{IEEEeqnarray*}
For a check node with degree $r$ and rank $r$, with probability
$r/R_0(t)$ it is $U$, and hence connects to $V$, otherwise, with
probability $r/(K-t)$ it connects to $V$. Therefore, 
\begin{equation*}
  N_{r,r} \sim \text{Binom}\left(\frac{R_{r,r}(t)}{r},
    \frac{r}{R_0(t)} + \left(1-\frac{r}{R_0(t)}\right)\frac{r}{K-t}\right).
\end{equation*}
 Taking expectation on \eqref{eq:trenddr0}, we have
\begin{IEEEeqnarray*}{rCl}
  \E[R_0(t+1) - R_0(t)|\bar R(t)] 
  & = & 
  \sum_{r} r \alpha_{r+1,r} \frac{R_{r+1,r}(t)}{K-t} - \sum_r \left(\frac{R_{r,r}(t)}{R_0(t)}+
  \left(1-\frac{r}{R_0(t)}\right)\frac{R_{r,r}(t)}{K-t}\right) \\
  & = & 
  \sum_{r} r \alpha_{r+1,r} \frac{R_{r+1,r}(t)}{K-t} - \frac{R_{0}(t)}{K-t} - 1 + \sum_r 
  \frac{r}{R_0(t)}\frac{R_{r,r}(t)}{K-t}. 
 \end{IEEEeqnarray*}
 The expectation in \eqref{eq:d99c} is obtained by noting that $\sum_r 
  \frac{r}{R_0(t)}\frac{R_{r,r}(t)}{K-t} < \frac{M^2}{K(1-c)}$ since $t\leq cK$.
\end{IEEEproof}

\subsection{Sufficient and Necessary Conditions}

We care about when $R_0(t)$ goes to zero for the first time. The
evolution of $R_0(t)$ depends on that of $R_{d,r}(t)$, $(d,r)\in
\mathcal{F}$. To study the trend, we can approximate 
$R_{d,r}(t)$ by its expectation. 
Consider the system of differential equations
\begin{IEEEeqnarray}{rCl}
  \frac{\diff\rho_{d,r}(\tau)}{\diff \tau}
  & = & \big(\alpha_{d+1,r}\rho_{d+1,r}(\tau)+\bar \alpha_{d+1,r+1}\rho_{d+1,r+1}(\tau)  -\rho_{d,r}(\tau)\big) \frac{d}{\theta -\tau}, \quad (d,r)\in \mathcal{F},  \label{eq:df1} \\
  \frac{\diff \rho_0(\tau)}{\diff \tau} 
  & = & \frac{\sum_{r=1}^{D-1}r\alpha_{r+1,r}
    \rho_{r+1,r}(\tau) -  \rho_0(\tau)}{\theta -\tau} - 1 \label{eq:df2}
\end{IEEEeqnarray}
with initial values $\rho_{d,r}(0) = \rho_{d,r}$, $(d,r)\in
\mathcal{F}$, and $\rho_0(0) = \sum_r \rho_{r,r}$, where $\theta =
K/n$ is the design rate of the BATS code.

We can get some intuition about how the system of differential
equations is obtained by substituting $R_{d,r}(t)$ and $ R_0(t)$ with
$n\rho_{d,r}(t/n)$ and $n \rho_0(t/n)$, respectively, in \eqref{eq:d9}
and \eqref{eq:d99c}. Defining $\tau=t/n$ and letting $n\rightarrow
\infty$, we obtain the system of differential equations in
\eqref{eq:df1} and \eqref{eq:df2}.  The expectation is ignored because
$\rho_{d,r}(\tau)$ and $ \rho_0(\tau)$ are deterministic
functions. Theorem~\ref{the:wormald} in Section~\ref{sec:prove} makes the
above intuition rigorous.

The system of differential equations in \eqref{eq:df1} and
\eqref{eq:df2} is solved in Appendix~\ref{sec:sol} for $0\leq \tau <
\theta$.  %
In particular, the solution for $\rho_0(\tau)$ is
\begin{IEEEeqnarray*}{rCl}
    \rho_0(\tau) & = & \left(1-\frac{\tau}{\theta}\right) \Bigg(\sum_{r = 1}^M  \alpha_{r+1,r} \sum_{d = r+1}^D \rho_{d,r}^{(d-r-1)}  \Bi_{d-r,r}\left(\frac{\tau}{\theta}\right) %
    + \sum_{r = 1}^M \rho_{r,r} + \theta \ln(1-\tau/\theta)
    \Bigg), \IEEEyesnumber \label{eq:solr0}
\end{IEEEeqnarray*}
where $ \rho_{d,r}^{(d-r-1)}$ is defined by the recursive formula
\begin{IEEEeqnarray}{rCl}
   \rho_{d,r}^{(0)} & \triangleq  & \rho_{d,r},  \label{eq:rdr0} \\
   \rho_{d,r}^{(i+1)} & \triangleq & \alpha_{d-i,r} \rho_{d,r}^{(i)} + \bar \alpha_{d-i,r+1}  \rho_{d,r+1}^{(i)}; \label{eq:rdr1}
\end{IEEEeqnarray}
and
\begin{align*}
  \Bi_{a,b}(x) \triangleq \sum_{j=a}^{a+b-1}\binom{a+b-1}{j} x^j(1-x)^{a+b-1-j}
\end{align*}
is called the \emph{regularized incomplete beta function}.  For $\bar \eta\in
(0,1)$, the following theorem shows that if $\rho_0(\tau)>0$ for
$\tau \in [0, \bar\eta]$, then the
decoding does not stop until $t> \bar\eta K$  with high probability, and $R_{d,r}(t)$ and $
R_0(t)$ can be approximated by $n\rho_{d,r}(t/n)$ and $n \rho_0(t/n)$,
respectively.

\begin{theorem} \label{the:1} Consider a sequence of decoding graphs
  $\text{BATS}(K, n, \Psi, h)$, $n=1,2,\ldots$ with fixed $\theta = K/n$,
  and the empirical rank distribution of transfer matrices
  $(\pi_0,\ldots,\pi_M)$ satisfying
  \begin{equation}\label{eq:sssc}
    |\pi_i - h_i| = \bigO(n^{-1/6}), \quad 0\leq i \leq M,
  \end{equation}
  with probability at least $1 - \gamma(n)$, where $\gamma(n) = o(1)$.
  For $\bar \eta\in (0,1)$,
  \begin{itemize}
  \item[(i)] if $\rho_0(\tau)>0$ for $\tau\in[0, \bar\eta\theta]$, then
  for sufficiently large $K$,
  with probability $1 - \bigO(n^{7/24}\exp(-n^{1/8})) -
  \gamma(n)$, the decoding terminates with at least $\bar \eta K$
  variable nodes decoded, and 
  \begin{IEEEeqnarray*}{rCl}
    |R_{d,r}(t) - n \rho_{d,r}(t/n)| & = & \bigO(n^{5/6}), \ (d,r)\in \mathcal{F} \\
    | R_0(t) - n  \rho_0(t/n) | & = & \bigO(n^{5/6})
  \end{IEEEeqnarray*}
  uniformly for $t \in [0, \bar\eta K]$;
  \item[(ii)] if $\rho_0(\tau)<0$ for some $\tau\in[0, \bar\eta\theta]$, then
  for sufficiently large $K$,
  with probability $1 - \bigO(n^{7/24}\exp(-n^{1/8})) -
  \gamma(n)$, the decoding terminates before $\bar\eta K$
  variable nodes are decoded.
  \end{itemize}
\end{theorem}

\bigskip

The quantities defined in \eqref{eq:rdr0} and \eqref{eq:rdr1} warrant
some interpretation. Consider a totally random $d\times M$ matrix $G_d$
and transfer matrix $H$ with $\rank(H)$ following
the distribution of the probability vector $h$.
Let $G_d^{(0)}$ be $G$ and for $t>0$, $G_d^{(t)}$ be the submatrix of
$G_d^{(t-1)}$ with a row deleted.
We show that
\begin{equation} \label{eq:eqq}
  \rho_{d,r}^{(t)}  =  d \Psi_d \Pr\{\rank(G_d^{(t)}H)=r\}.
\end{equation}
By \eqref{eq:cp1}, we have
\begin{equation*}
  \rho_{d,r}^{(0)} = \rho_{d,r} = d \Psi_d \Pr\{\rank(G_dH)=r \}.
\end{equation*}
Assume that \eqref{eq:eqq} holds for all time up to $t-1$. 
By definition,
\begin{IEEEeqnarray*}{rCl}
  \rho_{d,r}^{(t)} 
  & = & 
  \alpha_{d-t+1,r} \rho_{d,r}^{(t-1)}  + \bar
  \alpha_{d-t+1,r+1} \rho_{d,r+1}^{(t-1)} \\
  & = &  
  \Pr\{\rank(G_d^{(t)}H)=r | \rank(G_d^{(t-1)}H)=r\} d \Psi_d
  \Pr\{\rank(G_d^{(t-1)}H)=r\} \\
  & & + \IEEEyesnumber \label{eq:c8sdds}
  \Pr\{\rank(G_d^{(t)}H)=r | \rank(G_d^{(t-1)}H)=r+1\} d \Psi_d
  \Pr\{\rank(G_d^{(t-1)}H)=r+1\} \\
  & = & 
  d \Psi_d\Pr\{\rank(G_d^{(t)}H)=r\},
\end{IEEEeqnarray*}
where \eqref{eq:c8sdds} follows from Lemma~\ref{lemma:29s0} and the
induction hypothesis. 

In the expression of $\rho_0(\tau)$, we have 
\begin{IEEEeqnarray*}{rCl}
  \alpha_{r+1,r}\rho_{d,r}^{(d-r-1)} & = & d \Psi_d \alpha_{r+1,r} 
  \Pr\{\rank(G_d^{(d-r-1)}H)=r\} \\
  & = & 
  d \Psi_d \alpha_{r+1,r} \Pr\{\rank(G_{r+1}H)=r\} \\
  & = & 
  d \Psi_d \Pr\{\rank(G_{r+1}^{(1)}H) = \rank(G_{r+1}H)=r\}.
\end{IEEEeqnarray*}
Define the \emph{effective rank distribution} $
(\hbar_r=\hbar_r(h), r=1,\ldots, M)$ for a rank distribution 
$h$ as
\begin{equation}\label{eq:1}
  \hbar_r(h) = \Pr\{\rank(G_{r+1}^{(1)}H) = \rank(G_{r+1}H)=r\} = \sum_{i=r}^M \frac{\zeta_r^i}{q^{i-r}} h_i,
\end{equation}
which is the probability that a batch is
decodable for the first time when its degree becomes $r$.
Let
\begin{equation}\label{eq:2}
  \hbar_r' = \Pr\{\rank(G_{r}H)=r\} = \sum_{k=r}^M \zeta_r^k h_k,
\end{equation}
which is the
probability that a batch is decodable when its degree is $r$. 
We can write $\rho_{r,r} = r\Psi_r \hbar_r'$.
Using the above notations, we can simplify the expression of
$\rho_0(\tau)$ as 
\begin{IEEEeqnarray*}{rCl}
    \rho_0(\tau) & = & \left(1-\frac{\tau}{\theta}\right)
    \Bigg(\sum_{r = 1}^M \sum_{d = r+1}^D d\Psi_d \hbar_r  \Bi_{d-r,r}\left(\frac{\tau}{\theta}\right) 
    + \sum_{r = 1}^M r\Psi_r \hbar_r' + \theta \ln(1-\tau/\theta)
    \Bigg). \IEEEyesnumber \label{eq:ccis1s}
\end{IEEEeqnarray*}

Once a batch becomes decodable, it remains decodable until all its
contributors are decoded. This statement is equivalent to the
following lemma.
\begin{lemma} \label{lemma:4irw8}
  $\hbar_r' = \sum_{k\geq r} \hbar_k.$
\end{lemma}
\begin{IEEEproof}
  By definition,
  \begin{IEEEeqnarray*}{rCl}
    \hbar_r' & = & \Pr\{\rank(G_{r}H)=r\} \\
    & = & 
    \Pr\{\rank(G_{r+1}^{(1)}H)=r, \rank(G_{r+1}H)=r\} +
    \Pr\{\rank(G_{r+1}^{(1)}H)=r,\rank(G_{r+1}H)=r+1\} \\
    & = & 
    \hbar_r + \Pr\{\rank(G_{r+1}H)=r+1\}.
  \end{IEEEeqnarray*}
  The proof is completed by expanding the formula recursively.
\end{IEEEproof}

\section{Proof of Theorem~\ref{the:1}}
\label{sec:prove}

\subsection{A General Theorem}

The main technique to prove Theorem~\ref{the:1} is a general
theorem by Wormald \cite{wormald95,wormald99} with a small modification.
The statement of the next theorem follows that of
\cite[Theorem~5.1]{wormald99} with an extra initial condition.  A
similar version is provided in \cite[Theorem~C.28]{RU08} with a deterministic
boundedness condition.

We say a function $f(u_1,\ldots,u_j)$ satisfies a \emph{Lipschitz condition}
on $\mathcal{D}\subset \mathbb{R}^j$ if there exists a constant $C_L$ such that
\begin{equation*}
  |f(u_1,\cdots,u_j) - f(v_1,\cdots,v_j)| \leq C_L\max_{1\leq i \leq j}|u_i-v_i|
\end{equation*}
for all $(u_1,\cdots,u_j)$ and $(v_1,\cdots,v_j)$ in $\mathcal{D}$. 
We call $C_L$ the Lipschitz constant for $f$.
Note that $\max_{1\leq i \leq j}|u_i-v_i|$ is the distance between 
$(u_1,\cdots,u_j)$ and $(v_1,\cdots,v_j)$ in the $l^\infty$-norm.

\begin{theorem}\label{the:wormald}
  Let $\mathcal{G}_0,\mathcal{G}_1,\ldots$ be a random process with a positive integer parameter $n$,
  and let $(Y_l(t))_{l=0}^L$ be a random vector
  determined by $\mathcal{G}_0,\ldots,\mathcal{G}_t$. For some constant $C_0$ and all
  $l$, $|Y_l(t)| < C_0n$ for $t\geq 0$ and all $n$.  Let $\mathcal{D}$
  be some bounded connected open set containing the closure of
  \begin{equation*}
    \{(0,z_1,\ldots,z_L):\exists n, \Pr\{Y_l(0)=z_l n, 1\leq l \leq L\} \neq 0\}.
  \end{equation*}
  Define the \emph{stopping time} $T_{\mathcal{D}}$ to be the minimum
  $t$ such that $(t/n,Y_1(t)/n,\ldots,Y_L(t)/n)\notin \mathcal{D}$.
  Assume the following conditions hold.
  \begin{itemize}
  \item[(i)] (Boundedness) 
    For some functions $\beta = \beta(n) \geq 1$ and $\gamma=\gamma(n)$,
    the probability that 
    \begin{equation*}
      \max_l|Y_l(t+1)-Y_l(t)|\leq \beta,
    \end{equation*}
    is at least $1-\gamma$
    for $t<T_{\mathcal{D}}$.
  \item[(ii)] (Trend) For some function $\lambda_1 = \lambda_1(n)=o(1)$,
    if $t<T_{\mathcal{D}}$,
    \begin{IEEEeqnarray*}{rCl}
      \E[Y_l(t+1)-Y_l(t)| \mathcal{G}_1,\ldots,\mathcal{G}_t]
        & = & f_l\left(\frac{t}{n}, \left(\frac{Y_{i}(t)}{n}\right)_{i=0}^L\right) + \bigO(\lambda_1),
    \end{IEEEeqnarray*}
    for $1\leq l\leq L$.
  \item[(iii)]  (Lipschitz)  Each function $f_l$ satisfies a
     Lipschitz condition on 
     $\mathcal{D} \cap \{(t,z_1,\ldots,z_L),t\geq 0\}$
    with the same Lipschitz constant $C_L$ for each $l$.
  \item[(iv)] (Initial condition) For some point $(0, z_1^0,\ldots,z_l^0)\in\mathcal{D}$,
    \begin{equation*} 
      |Y_l(0)/n - z_l^0| \leq \sigma = o(1), 0\leq l \leq L.
    \end{equation*}
  \end{itemize}
  Then the following are true. 
  \begin{itemize}
  \item[(a)]
    For $(0,(\hat{z}_l)_{l=1}^L) \in \mathcal{D}$, the system of
    differential equations
    \begin{equation*}
      \frac{\diff z_l(\tau)}{\diff \tau} = f_l(\tau,(z_{l'}(\tau))_{l'=1}^L), \quad l = 1,\ldots, L,
    \end{equation*}
    has a unique solution in $\mathcal{D}$ for
    $z_l:\mathbb{R}\rightarrow \mathbb{R}$ passing through
    $z_l(0)= \hat{z}_l$, $l = 1,\ldots, L$, and this solution extends
    to points arbitrarily close to the boundary of $\mathcal{D}$.
  \item[(b)] Let $\lambda >\max\{\sigma, \lambda_1+C_0n\gamma\}$
    with $\lambda = o(1)$. There exists a sufficiently large constant
    $C_1$ such that when $n$ is sufficiently large, with probability
    $1-\bigO(n\gamma+\frac{\beta}{\lambda} \exp(-\frac{n
      \lambda^3}{\beta^3}))$,
    \begin{equation}
      |Y_l(t) -  n z_l(t/n)| = \bigO(\lambda n) \label{eq:mru}
    \end{equation}
    uniformly for $0\leq t \leq \bar \tau n$ and for each $l$, where 
    $\hat z_l = z_l^0$, and
    $\bar \tau = \bar \tau(n)$ is the supremum of those
    $\tau$ to which the solution of the system of differential
    equations in (a) can be extended before reaching within
    $l^\infty$-distance $C_1\lambda$ of the boundary of $\mathcal{D}$.
  \end{itemize}
\end{theorem}

\begin{IEEEproof}
  The proof follows exactly the proof of \cite[Theorem~5.1]{wormald99}
  except for the place where we need to handle the initial condition (iv).  We
  only have to modify the definition of $B_j$ (below (5.9) in
  \cite{wormald99}) in the original proof to
  \begin{equation*}
    B_j =\left(n\lambda+{\omega}\right)\left( \left(1+\frac{B\omega}{n}\right)^j-1\right) + B_0\left(1+\frac{B\omega}{n}\right)^j,
  \end{equation*}
  where $B_0 = n\lambda$.  The induction in the original proof now
  begins by the fact that $|z_l(0) - Y_l(0)/n| \leq \sigma <
  \bigO(\lambda)$. The other part of the proof stays the same as that
  of \cite[Theorem~5.1]{wormald99}.
\end{IEEEproof}

\subsection{Completing the Proof}

We first prove two technical lemmas.
For $\text{BATS}(K, n, \Psi, h)$, the degrees of the variable nodes are
not independent but follow the same distribution.  The following lemma
shows that the degree of a variable node is not likely to be much larger than its
expectation.

\begin{lemma} \label{lemma:cca} Let $V$ be the degree of a variable
  node of $\text{BATS}(K, n, \Psi, h)$. For any $\alpha>0$,
  \begin{equation*}
  \Pr\{V > (1+\alpha)\bar{\Psi}/\theta \} < \left(\frac{e^{\alpha}}{(1+\alpha)^{(1+\alpha)}}\right)^{\bar{\Psi}/\theta},
  \end{equation*}
  where $\theta = K/n$.
\end{lemma}
\begin{IEEEproof}%
  Fix a variable node.  Let $X_i$ be the indicator random variable
  of the $i$th check node being the neighbor of the specific variable node. 
  Then $V = \sum_i X_i$.
  We have $\E[V]  = \sum_i \E[X_i] =  \sum_i \sum_d \frac{d}{K} \Psi_d
     =  \frac{n}{K}\bar{\Psi}  = \frac{\bar{\Psi}}{\theta}$.
  Since $X_i$, $i=1,\ldots,n$, are mutually independent, the lemma is proved by applying
  the Chernoff bound. %
\end{IEEEproof}

The following lemma verifies the boundedness condition of Theorem~\ref{the:wormald}.

\begin{lemma} \label{cor:boundness}
When $\beta/D> \bar{\Psi}/\theta$, the probability that 
\begin{equation*}
  \max_{\iota \in \mathcal{F}\cup \{0\}} |R_{\iota}(t+1) - R_{\iota}(t)| \leq \beta,
\end{equation*}
is at least 
\begin{equation*}
  1 - \theta n \exp\left(-\frac{\beta}{D}(\ln (\beta/D) - \ln (\bar{\Psi}/\theta)  - 1)- \frac{\bar{\Psi}}{\theta} \right).
\end{equation*}
\end{lemma}
\begin{IEEEproof}
  Let $V$ be the degree
of the variable node to be removed at the beginning of time $t+1 $. 
By \eqref{eq:trenddr}, we have for $(d,r)\in \mathcal{F}$,
\begin{equation*}
  |R_{d,r}(t+1) - R_{d,r}(t)| \leq DV,
\end{equation*}
  and by \eqref{eq:trenddr0}, we have 
  \begin{equation*}
    |R_0(t+1) - R_0(t)| \leq DV.
  \end{equation*}
Hence when $\beta/D> \bar{\Psi}/\theta$,
\begin{IEEEeqnarray*}{rCl}
  \IEEEeqnarraymulticol{3}{l}{\Pr\left\{ \max_{\iota \in \mathcal{F}\cup \{0\}} |R_{\iota}(t+1) - R_{\iota}(t)| \leq \beta \right\}} \\ \ 
  & \geq & \Pr\{V D \leq \beta \}  \\
  & \geq & \Pr\{\text{the degrees of all variable nodes at time zero} \leq \beta/D \} \\
  & > & 1 - \theta n\exp\left(-\frac{\beta}{D}(\ln (\beta/D) - \ln (\bar{\Psi}/\theta)  - 1)- \frac{\bar{\Psi}}{\theta} \right),
\end{IEEEeqnarray*}
where the last inequality follows from Lemma~\ref{lemma:cca} and the union bound.
\end{IEEEproof}

\begin{IEEEproof}[Proof of Theorem~\ref{the:1}]
  We consider in the proof only the instances of $\text{BATS}(K, n,
  \Psi, h)$ satisfying
  \begin{equation}\label{eq:typicadddsl}
    \left|\frac{{R}_{d,r}}{n} - \rho_{d,r}\right| = \bigO(n^{-1/6}), \quad (d,r)\in \bar{\mathcal{F}}.
  \end{equation}
  By Lemma~\ref{lemma:init} this will decrease the probability bounds
  we will obtained by at most $\gamma(n)+2MD\exp(-2n^{2/3})$.

  Define the stopping time $T_{0}$ as the first time $t$ such that
  $R_0(t)=0$.  By defining suitable functions $f_{d,r},
  (d,r)\in\mathcal{F}$ and $f_0$ we can rewrite
  \eqref{eq:d9} and \eqref{eq:d99c} as
\begin{IEEEeqnarray*}{rCl}
  \IEEEeqnarraymulticol{3}{l}{\E[R_{d,r}(t+1)-R_{d,r}(t)|\bar R(t)]} \\ \ 
  & = & f_{d,r}\left(\frac{t}{n}, \left(\frac{R_0(t)}{n}\right), \left(\frac{R_{d',r'}(t)}{n}\right)_{(d',r')\in\mathcal{F}}\right),\  (d,r)\in\mathcal{F}  \\
  \IEEEeqnarraymulticol{3}{l}{\E[ R_0(t+1) - R_0(t)|\bar R(t)]}  \\ \
  & = & f_{0}\left(\frac{t}{n}, \left(\frac{R_0(t)}{n}\right), \left(\frac{R_{d',r'}(t)}{n}\right)_{(d',r')\in\mathcal{F}}\right) + \bigO\left(\frac{1}{n}\right),
\end{IEEEeqnarray*}
for $t<T_0$.  For $\iota \in \mathcal{F}\cup\{0\}$, define random
variable $\hat{R}_{\iota}$ as $\hat{R}_{\iota}(0) = R_{\iota}(0)$ and
for $t\geq 0$,
  \begin{IEEEeqnarray*}{rCl}
  \hat{R}_{\iota}(t+1)
   & = &  \left\{ 
       \begin{array}{ll}
         R_{\iota}(t+1) & t<T_0 \\
         \hat{R}_{\iota}(t) 
          + 
     f_{\iota}\left(\frac{t}{n}, \left(\frac{R_0(t)}{n}\right),
    \left(\frac{R_{d',r'}(t)}{n}\right)_{(d',r')\in\mathcal{F}}\right) & t\geq T_0.
       \end{array}
       \right.
  \end{IEEEeqnarray*}
  Note that $T_0$ is also the first time that
  $\hat{R}_0(t)$ becomes zeros.

  We now apply Theorem~\ref{the:wormald} with
  $(\hat{R}_0(t),(\hat{R}_{d,r}(t))_{(d,r)\in\mathcal{F}})$ in place of
  $(Y_l(t))_{l=1}^L$. The region $\mathcal{D}$ is defined as
  \begin{equation*}
    \mathcal{D} = (-\eta, (1-\eta/2)\theta) \times (-M,
  M+\eta) \times (-\eta,d)^{|\mathcal{F}|}.
  \end{equation*}
  So 1) $t/n$ is in
  the interval $(-\eta, (1-\eta/2)\theta)$; 2)
   $\hat R_{0}(t)/n$ is in the interval $(-M,
  M+\eta)$; and 3) $\hat R_{d,r}(t)/n$, 
  $(d,r)\in\mathcal{F}$, is in the interval $(-\eta,d)$.  As required,
  $\mathcal{D}$ is a bounded connected open set and containing all the
  possible initial state
  $(0,\hat{R}_{0}(0)/n,(\hat{R}_{d,r}(0)/n)_{(d,r)\in\mathcal{F}})$.

  The conditions of Theorem \ref{the:wormald} can readily be
  verified. When $t\geq T_0$, the change $|\hat{R}_{\iota}(t+1) -
  \hat{R}_{\iota}(t)|$ for $\iota \in \mathcal{F}\cup\{0\}$ is
  deterministic and upper bounded. When $t<T_0$, by
  Lemma~\ref{cor:boundness} with $\beta = n^{1/8}$, the
  boundedness condition (i) holds with
\begin{equation*}
  \gamma =  n \exp\left(-  n^{1/8} \left(c_{1,3}\ln n - c_{1,1}\right)- c_{1,2} \right),
\end{equation*}
where $c_{1,1}$, $c_{1,2}$, and $c_{1,3}$ are only related to $\bar{\Psi}$ and $\theta$.
The trend condition (ii) is satisfied with $\lambda_1 = \bigO(1/n)$.
By definition, it can be verified that $f_{\iota}$, $\iota\in
\mathcal{F}\cup\{0\}$ satisfy the Lipschitz condition (iii).
The initial condition (iv) holds with $\sigma = \bigO(n^{-1/6})$.

Wormald's method leads us to consider the system of
 differential equations
\begin{IEEEeqnarray*}{rCl}
  \frac{\diff \rho_{d,r}(\tau)}{\diff \tau} & = & f_{d,r}(\tau,\rho_0(\tau), (\rho_{d',r'}(\tau))_{(d',r')\in\mathcal{F}}), \quad  (d,r)\in\mathcal{F}\\
  \frac{\diff \rho_0(\tau)}{\diff \tau} & = & f_{0}(\tau, \rho_0(\tau), (\rho_{d',r'}(\tau))_{(d',r')\in\mathcal{F}})
\end{IEEEeqnarray*}
with the initial condition $\rho_{d,r}(0) = \rho_{d,r}$, $(d,r)\in\mathcal{F}$, and
$\rho_{0}(0) = \sum_{r}\rho_{r,r}$.  The conclusion (a) of
Theorem~\ref{the:wormald} shows the existence and uniqueness of the
solution of the above system of differential equations. We solve 
the system of differential equations explicitly in Appendix~\ref{sec:sol}.

Let $\lambda = \bigO(n^{-1/6})$. By the conclusion (b) of
Theorem~\ref{the:wormald}, we know that for a sufficiently large
constant $C_1$, with probability $1-\bigO(n\gamma +
\frac{\beta}{\lambda}\exp(-\frac{n \lambda^3}{\beta^3}))$,
\begin{IEEEeqnarray*}{rCl}
    |\hat{R}_{d,r}(t) - n \rho_{d,r}(t/n)| & = & \bigO(n^{5/6}),\ (d,r)\in \mathcal{F}, \\
    |\hat{R}_0(t)- n \rho_0(t/n)| & = & \bigO(n^{5/6})
\end{IEEEeqnarray*}
uniformly for $0\leq t \leq \bar \tau n$, where $\bar \tau$ is defined in Theorem~\ref{the:wormald}.
Increase $n$ if necessary so that $\frac{\beta}{\lambda}\exp(-\frac{n \lambda^3}{\beta^3}) =
n^{7/24}\exp(-n^{-1/8}) > n\gamma$  and $C_1\lambda < \frac{\eta}{2}\theta$, which implies $\bar \tau \geq (1-\eta)\theta$.
So there exists constants $c_{0}$ and $c_{0}'$ such that the event 
\begin{equation*}
 E_0 = \{|\hat{R}_0(t)/n - \rho_0(t/n)| \leq c_{0}n^{-1/6},\ 0\leq t \leq (1-\eta)K\}
\end{equation*}
holds with probability at least $1- c_0' n^{7/24}\exp(-n^{-1/8})$. 

Now we consider the two cases in the theorem to prove. (i) If $\rho_0(\tau)>0$ for $\tau \in [0, (1-\eta)\theta]$, then there exists $\epsilon>0$ such that $\rho_0(\tau)\geq \epsilon$ for $\tau \in [0, (1-\eta)\theta]$.
Increase $n$ if necessary so that $c_0n^{-1/6} < \epsilon$. Then, we have
\begin{IEEEeqnarray*}{rCl}
  \Pr\{T_0>(1-\eta)K\} & = & \Pr\{\hat{R}(t)>0, 0\leq t \leq (1-\eta)K \}\\
  & \geq & \Pr\{E_0\} \IEEEyesnumber \label{eq:ccsz} \\
  & \geq & 1- c_0' n^{7/24}\exp(-n^{-1/8}),
\end{IEEEeqnarray*}
where \eqref{eq:ccsz} follows that under the condition $E_0$, for
all $t\in[0, (1-\eta)K]$, $\hat{R}_0(t)/n \geq \rho_0(t/n) -
c_0n^{-1/6} > 0$.  Since $\hat{R}_{\iota} = R_{\iota}$, $\iota\in
\mathcal{F}\cup\{0\}$, when $t<T_0$, the first part of the theorem is
proved.

(ii) Consider $\rho_0(\tau_0)<0$ for $\tau_0 \in [0,
(1-\eta)\theta]$. There exists $\epsilon>0$ such that
$\rho_0(\tau) \leq -\epsilon$ for all $\tau \in [\tau_0-\epsilon, \tau_0+\epsilon]\cap
[0,(1-\eta)\theta]$.
Increase $n$ if necessary so that $c_0n^{-1/6} < \epsilon$ and $n\epsilon >1$.
Then, we have
\begin{IEEEeqnarray*}{rCl}
  \IEEEeqnarraymulticol{3}{l}{\Pr\{T_0 \leq (1-\eta)K\}} \\ \quad
  & = & \Pr\{\hat{R}_0(t)<0,\ \text{for some}\ t\in [0, (1-\eta)K] \} \\
  & \geq & \Pr\{E_0\} \IEEEyesnumber \label{eq:ccsa} \\
  & \geq & 1- c_0' n^{7/24}\exp(-n^{-1/8}),
\end{IEEEeqnarray*}
where \eqref{eq:ccsa} can be shown as follows. Since $n\epsilon>1$, there exists
$t_0$ such that $t_0/n \in [\tau_0-\epsilon, \tau_0+\epsilon]\cap
[0,(1-\eta)\theta]$. Hence, under the condition $E_0$,
$\hat{R}_0(t_0)/n \leq c_0n^{-1/6} + \rho_0(t_0/n) < 0$.  

The proof of the theorem is completed by subtracting the probability that \eqref{eq:typicadddsl} does not hold.
\end{IEEEproof}

\section{Degree Distribution Optimizations}
\label{sec:deg}

Theorem~\ref{the:1} gives a sufficient condition 
that the BP decoding succeeds with high probability. This condition
induces optimization problems that generate degree distributions
meeting our requirement.

\subsection{Optimization for single rank distribution}

Let $\hbar=(\hbar_1,\ldots,\hbar_M)$, where $\hbar_i$ is defined in \eqref{eq:1}.
Define
\begin{equation}\label{eq:omega}
  \Omega(x; \hbar,\Psi)  \triangleq  \sum_{r=1}^M \hbar_r
  \sum_{d=r+1}^D d\Psi_d \Bi_{d-r,r}(x) + \sum_{r=1}^M
  r\Psi_r\sum_{s = r}^M \hbar_s.
\end{equation}
When the context is clear, we also write $\Omega(x;\Psi)$,
$\Omega(x; \hbar)$ or $\Omega(x)$ to simplify the notation.
By \eqref{eq:ccis1s} and Lemma~\ref{lemma:4irw8}, we can write 
\begin{equation}\label{eq:cce}
 \rho_0(\tau) = (1-\tau/\theta)\left(\Omega(\tau/\theta) + \theta\ln(1-\tau/\theta) \right).
\end{equation}

For $\bar \eta\in (0,1)$, we say a rate $R$ is \emph{$\bar
  \eta$-achievable} by BATS codes using BP decoding if for every $\epsilon>0$ and every
sufficiently large $K$, there exists a BATS code with $K$ input
packets such that for $n\leq \bar\eta K/(R-\epsilon)$ received batches,
the BP decoding recovers at least $\bar \eta K$ input packets with
probability at least $1-\epsilon$.  Define the optimization problem
\begin{equation}\label{eq:op1}
  \max \theta   \quad \text{s.t.}  \left\lbrace
    \begin{IEEEeqnarraybox}[][c]{l}
      \Omega(x; \hbar(h),\Psi) + \theta \ln(1-x)\geq 0, \quad 0\leq x \leq \bar \eta,  \\
      \sum_d \Psi_d = 1 \ \ \text{and} \ \ \Psi_d \geq 0, \ d =
      1,\cdots, D.
    \end{IEEEeqnarraybox}
  \right. \tag{P1}
\end{equation}

\begin{lemma}\label{prop:1}
  Let $\hat \theta$ be the optimal value in \eqref{eq:op1}.  When the
  empirical rank distribution of the transfer matrices converges in
  probability to $h=(h_0,\ldots,h_M)$ (in the sense of
  \eqref{eq:sssc}), any rate less than or equal to $\bar \eta \hat
  \theta$ is $\bar \eta$-achievable by BATS codes using BP decoding.
\end{lemma}
\begin{IEEEproof}
  To show that $\bar\eta\hat \theta$ is $\bar \eta$-achievable, by Theorem~\ref{the:1}, we only
  need to show that there exists a degree distribution such that for any $\epsilon>0$,
  \begin{equation}
    \label{eq:cond1}
    \Omega(x) + (\hat \theta-\epsilon) \ln(1-x)> 0, \quad 0\leq x \leq \bar \eta.
  \end{equation}
  For the degree distribution $\Psi$ that
  achieves $\hat \theta$ in \eqref{eq:op1},
  we have 
  \begin{equation*}
    \Omega(x;\Psi) + \hat \theta\ln(1-x)\geq 0, \quad 0\leq x \leq \bar \eta.
  \end{equation*}
  Multiplying by $\frac{\hat \theta-\epsilon}{\hat \theta}$ on both
  sides, we have
  \begin{equation}
    \label{eq:cond1-1}
    \frac{\hat \theta-\epsilon}{\hat
    \theta}\Omega(x;\Psi) + (\hat \theta-\epsilon) \ln(1-x)\geq 0, \quad 0\leq x \leq \bar \eta.
  \end{equation}
  Since $\Omega(x;\Psi)>0$ for $x>0$, \eqref{eq:cond1-1} implies that
  $\Psi$ satisfies \eqref{eq:cond1} except possibly for $x=0$. 
  Checking the definition of $\Omega$ in \eqref{eq:omega}, we have
  $\Omega(0;\Psi)=\sum_{r=1}^M r\Psi_r\hbar_r'$.  If $\sum_{r=1}^M
  r\Psi_r\hbar_r'>0$, which implies $\Psi$ satisfies \eqref{eq:cond1}, we are
  done. In the following, we consider the case with $\sum_{r=1}^M r\Psi_r\hbar_r'=0$.

  Let $r^*$ be the largest integer $r$ such that $h_{r}>0$. It can be
  verified in \eqref{eq:2} that $\hbar_r'=0$ for $r>r^*$ and
  $\hbar_r'>0$ for $r\leq r^*$.  Since $\sum_{r=1}^M
  \hbar_r'r\Psi_r=0$, we know that $\sum_{d\leq r^*}\Psi_d=0$.  Define
  a new degree distribution $\Psi'$ by $\Psi_d'=\Psi_d \frac{\hat
    \theta-\epsilon}{\hat \theta}$ for $d>r^*$ and $\Psi_d' = \Delta$
  for $d\leq r^*$, where $\Delta> 0$ can be determined by the
  constraint $\sum_d\Psi'_d=1$.  Then we can check that $\Psi'$
  satisfies \eqref{eq:cond1}.
\end{IEEEproof}

The converse of Lemma~\ref{prop:1} is that ``a rate larger than $\hat \theta$ is not $\bar \eta$-achievable''.
Intuitively, for any $\epsilon>0$, we cannot have a degree distribution such that 
\begin{equation*}
    \Omega(x) + (\hat \theta+\epsilon) \ln(1-x)\geq 0, \quad 0\leq x \leq \bar \eta,
\end{equation*}
where $\hat\theta$ is the maxima of \eqref{eq:op1}.  Thus, with $\hat
\theta+\epsilon$ in place of $\theta$ in the expression of $\rho_0$ in
\eqref{eq:cce}, for any degree distribution we have $\rho_0(\tau)<0$
for some $\tau \in [0,\bar\eta (\hat \theta+\epsilon)]$.  By
Theorem~\ref{the:1}, for any degree distribution there exists $K_0$
such that when the number of input packets $K\geq K_0$, with
probability approaching 1 the BATS code cannot recover $\bar \eta K$
input packets. To prove this converse, however, we need a uniform
bound $K_0$ for all degree distributions such that the second part of
Theorem~\ref{the:1} holds, which is difficult to obtain. Instead, we
demonstrate that $\hat \theta$ is close to the capacity of the
underlying linear operator channel (cf. Section~\ref{sec:trans}).

Before analyzing the achievable rate, we determine the maximum degree
$D$, which affects the encoding/decoding
complexity. In Section~\ref{sec:rd}, we have assumed that $D=\bigO(M)$.
The next theorem shows that it is optimal to choose $D
= \lceil M/\eta\rceil - 1$, where $\eta = 1-\bar \eta$.

\begin{theorem}\label{the:D}
  Using $D>\lceil M/\eta\rceil - 1$ does not give a better optimal
  value in \eqref{eq:op1}, where $\eta = 1-\bar \eta$.
\end{theorem}
\begin{IEEEproof}
  Consider an integer $\Delta$ such that $\eta \geq
  \frac{M}{\Delta+1}$.  Let $\Psi$ be a degree distribution with
  $\sum_{d>\Delta} \Psi_d >0$. Construct a new degree distribution
  $\tilde \Psi$ as follows:
  \begin{align*}
  \tilde \Psi_d & = \Psi_d,\quad d<\Delta, \\
  \tilde \Psi_\Delta & = \sum_{d\geq \Delta} \Psi_d,\\
  \tilde \Psi_d &  = 0,\quad d>\Delta.
\end{align*}
We now show that $\Omega(x; \tilde \Psi) > \Omega(x; \Psi)$ for all
$0< x \leq 1-\eta$.  Write
\begin{IEEEeqnarray*}{rCl}
 \IEEEeqnarraymulticol{3}{l}{\Omega(x; \tilde \Psi) - \Omega(x; \Psi)} \\\quad
 & = & \sum_{d=\Delta+1}^{\infty} \Psi_d \sum_{r=1}^M \hbar_r (\Delta \Bi_{\Delta-r,r}(x) - d \Bi_{d-r,r}(x)).
\end{IEEEeqnarray*}
For $d\geq \Delta+1$, 
\begin{align*}
  \frac{r-1}{d-r} \leq \frac{M-1}{d-M} < \frac{M}{\Delta-M+1} \leq \frac{\eta}{1-\eta}.
\end{align*}
So we can apply the properties of the incomplete beta function (Lemma~\ref{lemma:2c} in Appendix~\ref{sec:beta}) to
show that, for any $x$ with $0<x\leq 1-\eta$,
\begin{align*}
 \frac{d \Bi_{d-r,r}(x)}{(d-1) \Bi_{d-1-r,r}(x)} & < \frac{d}{d-1}\left(1-\frac{\eta}{r}\right) \\
  & \leq \frac{d}{d-1}\left(1-\frac{\eta}{M}\right) \\
  & \leq \frac{\Delta+1}{\Delta} \left(1-\frac{1}{\Delta+1}\right) \\
  & = 1,
\end{align*}
which gives $\Omega(x; \tilde \Psi) > \Omega(x; \Psi)$ for $0< x \leq
1-\eta$.

Thus, for certain $\theta$ such that
\begin{equation*}
  \Omega(x; \Psi) + \theta \ln(1-x) \geq 0, \quad 0\leq x \leq 1-\eta,
\end{equation*}
we have %
\begin{IEEEeqnarray*}{rCl}
  \Omega(x;\tilde \Psi) + \theta \ln(1-x)  
   \geq 0, \quad 0\leq x \leq 1-\eta.
\end{IEEEeqnarray*}
This means that using only degree distributions $\Psi$ with
$\sum_{d>\Delta} \Psi_d = 0$, we can get the same optimal value as
using all degree distributions.  Therefore, it is sufficient to take the
maximum degree $D\leq \min_{\eta \geq \frac{M}{\Delta+1}} \Delta
=\lceil M/\eta\rceil -1$.
\end{IEEEproof}

To solve \eqref{eq:op1} numerically, one way is to relax it as a linear
programming by only considering $x$ in a linearly sampled set of values
between 0 and $1-\eta$. Let $x_i =
(1-\eta)\frac{i}{N}$ for some integer $N$.  We relax \eqref{eq:op1}
by considering only $x=x_i$, $i= 1,\ldots,N$, where $N$ can be chosen
to be $100$ or even smaller.

For many cases, we can directly use the degree distribution $\Psi$
obtained by solving \eqref{eq:op1} when block length $K$ is large. (We will discuss how to
optimize the degree distribution for small $K$ in
Section~\ref{sec:finite}.) 
But it is possible that $\Omega(0;\Psi)=0$, so that the
degree distribution $\Psi$ does not guarantee that decoding can
start.  We can then modify $\Psi$ as we do in
the proof of Lemma~\ref{prop:1} by increasing the probability
masses $\Psi_d$, $d\leq M$ by a small amount to make sure that decoding can
start.

\subsection{Achievable Rates}
\label{sec:achieve}

The first upper bound on the optimal value $\hat\theta$ of
\eqref{eq:op1} is given by the capacity of LOCs
with receiver side channel state information.  When the empirical rank
distribution of the transfer matrices converging to
$h=(h_0,\ldots,h_M)$, the capacity is $\sum_r rh_r$ packets per batch.
The BP decoding algorithm recovers at least a fraction $\bar \eta$
of all the input packets with high probability. So asymptotically BATS
codes under BP decoding can recover at least a fraction $\bar\eta\hat\theta$
of the input packets. Thus, we have $\bar \eta\hat\theta \leq
\sum_r rh_r$.

A tighter upper bound can be obtained by analyzing \eqref{eq:op1} directly.
Rewrite
\begin{IEEEeqnarray*}{rCl}
  \Omega(x;\Psi) & = &  \sum_{r=1}^M \hbar_r \sum_{d=r+1}^D d\Psi_d \Bi_{d-r,r}(x) + \sum_{r=1}^M \hbar_r \sum_{d=1}^r d\Psi_d, \\
  & = & \sum_{r=1}^M \hbar_r S_r(x;\Psi), \IEEEyesnumber \label{eq:omega2}
\end{IEEEeqnarray*}
where
\begin{equation}\label{eq:sr}
  S_r(x;\Psi) = S_r(x) \triangleq \sum_{d=r+1}^D d\Psi_d \Bi_{d-r,r}(x)+ \sum_{d=1}^r d\Psi_d. 
\end{equation}
This form of $\Omega(x;\Psi)$ will be used in the subsequent proofs.

\begin{theorem}\label{lemma:ub2}
The optimal value $\hat\theta$ of \eqref{eq:op1} satisfies
  \begin{equation*}
    \bar \eta \hat\theta \leq \sum_{r=1}^M r \hbar_r.
  \end{equation*}
\end{theorem}
\begin{IEEEproof}
  Fix a degree distribution that achieves the optimal value of \eqref{eq:op1}.
  Using \eqref{eq:dds} in Appendix~\ref{sec:beta}, we have
  \begin{IEEEeqnarray*}{rCl}
    \int_0^1 S_r (x)  \diff x & = & \sum_{d=r+1}^D d\Psi_d \int_0^1\Bi_{d-r,r}(x) \diff x  + \sum_{d=1}^r d\Psi_d \\
    & = & \sum_{d=r+1}^D r\Psi_d + \sum_{d=1}^r d\Psi_d \\
    & \leq & r \sum_{d=1}^D \Psi_d \\
    & = & r.
  \end{IEEEeqnarray*}
  Hence,
  \begin{equation}\label{eq:iocw1}
    \int_0^1\Omega(x)  \diff x 
     =  \int_0^1 \sum_{r=1}^M \hbar_r S_r(x) \diff x 
     \leq  \sum_{r=1}^M r \hbar_r.
  \end{equation}
  Since $\Omega(x)$ is an
  increasing function,
  \begin{equation}\label{eq:iocw2}
    \int_{1-\eta}^1 \Omega(x) \diff x \geq \eta \Omega(1-\eta) \geq - \eta \hat \theta \ln \eta.
  \end{equation}
  Since $\Omega(x) + \hat\theta \ln(1-x) \geq 0$ for $0<x\leq 1-\eta$,
  \begin{IEEEeqnarray*}{rCl}
    \int_0^{1-\eta} \Omega(x) \diff x  -
    \hat\theta(\eta\ln\eta + 1-\eta)
    & = & \int_0^{1-\eta} \Omega(x) \diff x + \hat\theta\int_0^{1-\eta} \ln(1-x) \diff x 
     \geq  0. \IEEEyesnumber \label{eq:iocw3}
  \end{IEEEeqnarray*}
  Therefore, by \eqref{eq:iocw1}-\eqref{eq:iocw3}, we have
  \begin{IEEEeqnarray*}{rCl}
    \sum_{r=1}^M r \hbar_r & \geq & \int_0^1\Omega(x)  \diff x \\
    & = & \int_0^{1-\eta} \Omega(x) \diff x + \int_{1-\eta}^1 \Omega(x) \diff x \\
    & \geq & \hat\theta(\eta\ln\eta + 1-\eta) - \eta \hat \theta \ln \eta \\
    & = & \hat\theta(1-\eta).
  \end{IEEEeqnarray*}
  The proof is completed.
\end{IEEEproof}

By Lemma~\ref{lemma:4irw8}, $\sum_{k=r}^M \hbar_{k} = \hbar_{r}' =
\sum_{i=r}^M h_i \cmatt{i}{r} \leq \sum_{k=r}^M h_k$, where the last
inequality follows from $\cmatt{i}{r}<1$.  Hence,
\begin{IEEEeqnarray*}{rCl}
  \sum_{r}r \hbar_r & = & \sum_{r=1}^M \sum_{k=r}^M \hbar_{k} \\
  & \leq & \sum_{r=1}^M \sum_{k=r}^M h_{k} \\
  & = & \sum_{r}r h_{r}.
\end{IEEEeqnarray*}
Therefore, 
Theorem~\ref{lemma:ub2} gives a strictly better upper bound than
$\sum_r rh_r$. When $q\rightarrow \infty$, $\sum_{r}r \hbar_r \rightarrow \sum_r r
h_{r}$. Even for small finite fields, $\sum_{r}r \hbar_r$ and $\sum_r r
h_{r}$ are very close.

We prove for a special case and demonstrate by simulation for
general cases that the optimal value $\hat\theta$ of
\eqref{eq:op1} is very close to $\sum_{r}r \hbar_r$.

\begin{theorem}\label{the:lower}
  The optimal value $\hat\theta$ of \eqref{eq:op1} satisfies
   $$\hat\theta \geq \max_{r=1,2,\cdots,M} r \sum_{i=r}^M \hbar_i.$$
\end{theorem}
\begin{IEEEproof}
Define a degree distribution $\Psi^r$ by
\begin{equation}\label{eq:phidr}
  \Psi_d^r = \left\{\begin{array}{ll} 
      0 & d\leq r, \\ 
      \frac{r}{d(d-1)} & d = r+1, \cdots, D-1, \\
      \frac{r}{D-1} & d = D.
           \end{array} \right.
\end{equation}
Recall the definition of $S_r(x;\Psi)$ in \eqref{eq:sr}.
For $M\geq r'\geq r$, we will show that 
\begin{equation}
  \label{eq:sii1}
  S_{r'}(x;\Psi^r) + r \ln(1-x) >0, \quad 0\leq x \leq 1 - \eta.
\end{equation}
By Lemma~\ref{lemma:cs} in Appendix~\ref{sec:beta},
\begin{align*}
  - r \ln(1-x) = r \sum_{d=r'+1}^{\infty} \frac{1}{d-1} \Bi_{d-r',r'}(x).
\end{align*}
By \eqref{eq:sr} and \eqref{eq:phidr},
\begin{IEEEeqnarray*}{rCl}
  S_{r'}(x;\Psi^r) + r\ln(1-x)
  & \geq & \sum_{d=r'+1}^D d\Psi_d^r \Bi_{d-r',r'}(x) -  r \sum_{d=r'+1}^{\infty} \frac{1}{d-1} \Bi_{d-r',r'}(x)\\
  & \geq & r\frac{D}{D-1}\Bi_{D-r',r'}(x) - r\sum_{d=D}^{\infty} \frac{1}{d-1} \Bi_{d-r',r'}(x) \\ 
  & = &  r\Bi_{D-r',r'}(x) - r\sum_{d=D+1}^{\infty} \frac{1}{d-1} \Bi_{d-r',r'}(x) .
\end{IEEEeqnarray*}
We will show that $\Bi_{D-r',r'}(x) > \sum_{d=D+1}^{\infty} \frac{1}{d-1}\Bi_{d-r',r'}(x)$ for $x\in [0, 1-\eta]$. This is equivalent to show that
\begin{align}\label{eq:d8s}
  \sum_{d=D+1}^{\infty} \frac{1}{d-1}\frac{\Bi_{d-r',r'}(x)}{\Bi_{D-r',r'}(x)} < 1 \quad \text{for}\ x\in [0, 1-\eta].
\end{align}
By Lemma~\ref{lemma:2d} in Appendix~\ref{sec:beta}, $\frac{\Bi_{d-r',r'}(x)}{\Bi_{D-r',r'}(x)}$ is monotonically increasing, so we only need to prove the above inequality for $x=1-\eta$. By Lemma~\ref{lemma:2c} in Appendix~\ref{sec:beta}, $\frac{\Bi_{d-r',r'}(1-\eta)}{\Bi_{D-r',r'}(1-\eta)} < (1-\frac{\eta}{M})^{d-D}$. Therefore,
\begin{align*}
  \sum_{d=D+1}^{\infty} \frac{1}{d-1}\frac{\Bi_{d-r',r'}(x)}{\Bi_{D-r',r'}(x)} & \leq  \frac{1}{D}\sum_{d=D+1}^{\infty} \frac{\Bi_{d-r',r'}(1-\eta)}{\Bi_{D-r',r'}(1-\eta)} \\ & 
	< \frac{1}{D}\sum_{d=D+1}^{\infty} (1-\frac{\eta}{M})^{d-D} \\ &  
	= \frac{M-{\eta}}{D{\eta}} \\ &
	\leq 1,
\end{align*}
where the last inequality follows from $D = \lceil M/\eta\rceil  -1$.
So we have established \eqref{eq:d8s} and hence \eqref{eq:sii1}.

Last, by \eqref{eq:omega2} and \eqref{eq:sii1}, we have for $0\leq x \leq 1 - \eta$,
\begin{align*}
  \Omega(x;\Psi^r) & \geq \sum_{r'\geq r} \hbar_{r'} S_{r'}(x;\Psi^{r}) \\ & 
  > -\ln(1-x)r\sum_{r'\geq r} \hbar_{r'},
\end{align*}
or
\begin{equation}
  \label{eq:ddbb}
  \Omega(x;\Psi^r) + \left(r\sum_{r'\geq r} \hbar_{r'}\right)\ln(1-x) > 0 .
\end{equation}
We conclude that $\hat\theta \geq r\sum_{r'\geq r} \hbar_{r'}$. 
The proof is completed
by considering all $r=1,2, \cdots, M$.
\end{IEEEproof}

Though in general the lower bound in Theorem~\ref{the:lower} is not
tight, we can show for a special case that it converges asymptotically
to the upper bound in Theorem~\ref{lemma:ub2}.  Consider a rank
distribution $h=(h_1,\ldots, h_M)$ with $h_{\kappa}=1$ for some $1\leq
\kappa \leq M$. Theorem~\ref{the:lower} implies that $\hat\theta \geq
\kappa \hbar_{\kappa}$. On the other hand, Theorem~\ref{lemma:ub2}
says that $(1-\eta)\hat\theta \leq \sum_{r}r \hbar_r = \kappa
\hbar_{\kappa} + \sum_{r< \kappa}r \hbar_r$. Note that $\eta$ can be
arbitrarily small, and $\sum_{r < \kappa}r \hbar_r \rightarrow 0$ and
$\hbar_{\kappa}\rightarrow h_{\kappa}$ when the field size goes to
infinity. Thus, both the upper bound in Theorem~\ref{lemma:ub2} and
the lower bound in Theorem~\ref{the:lower} converge to $\kappa
h_{\kappa}$, the capacity of the LOC with empirical rank distribution
converging to $h$.

\begin{figure}
  \centering

  \begin{tikzpicture}[scale=1]
    \begin{semilogyaxis}[xmin=0.88, xmax=1,no markers,
       ymin=0, ymax=1, legend pos=north west,ylabel=eCDF
      ]
       \addplot[blue] table {data/cdf1.txt};
       \addlegendentry{$\eta = 0.005$}
       \addplot[red,densely dashdotted] table {data/cdf2.txt};
       \addlegendentry{$\eta = 0.01$}
       \addplot[dashed] table {data/cdf3.txt};
       \addlegendentry{$\eta = 0.02$}
    \end{semilogyaxis}
  \end{tikzpicture}

  \caption{The empirical cumulative distribution function (eCDF) of
    $\tilde \theta := (1-\eta)\hat\theta/  \sum_{r} r \hbar_r$ for
    $4\times 10^4$ rank distributions. Here $q=2^8$ and $M=16$.}
  \label{fig:cdf}
\end{figure}
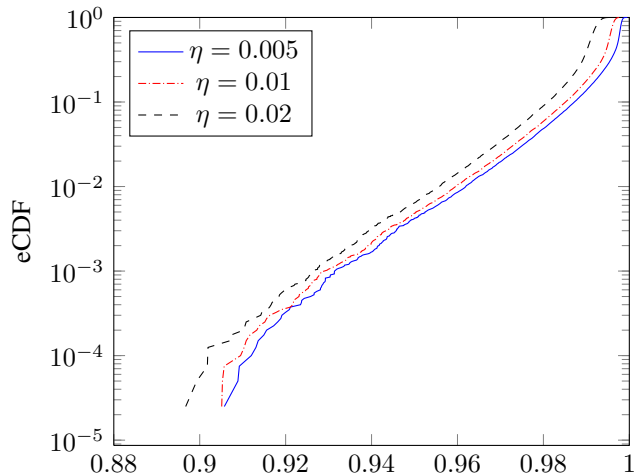

We can compute the achievable rates of BATS codes numerically by
solving \eqref{eq:op1}. Set
$M=16$ and $q=2^8$. Totally $4\times 10^4$ rank distributions are
tested\footnote{A rank distribution is randomly generated as
  follows. First, select $x_1,x_2,\ldots,x_{M-1}$ independently
  uniformly at random in $[0,\ 1]$. Next, sort $\{x_i\}$ such that
  $x_1\leq x_2 \leq \cdots \leq x_{M-1}$. Then, the rank distribution
  is given by $h_0=0$, for $1\leq r \leq M$, $h_r = x_r-x_{r-1}$,
  where $x_0=1$ and $x_M=1$. This gives an almost uniform sampling
  among all the rank distributions with $\sum_{i=1}^M h_i = 1$
  according to \cite{smith04}. The reason that we choose $h_0=0$ is as
  follows. For rank distribution $h=(h_0,\ldots,h_M)$ with $h_0>0$, we
  obtain a new rank distribution $h'=(h_0'=0,
  h_i'=h_i/(1-h_0),i=1,\ldots,M)$. Optimization \eqref{eq:op1} is
  equivalent for these two rank distributions except that the
  objective function is scaled by $1-h_0$. Thus the values of $\tilde
  \theta := (1-\eta)\hat\theta/ \sum_{r} r \hbar_r$ for both $h$ and
  $h'$ are the same.  }.  For each rank distribution $h$ we solve
\eqref{eq:op1} for $\eta=0.02, 0.01$ and $0.005$. The empirical
distributions of $\tilde \theta := (1-\eta)\hat\theta/ \sum_{r} r
\hbar_r$ are draw in Fig.~\ref{fig:cdf}. By Theorem~\ref{lemma:ub2},
$\tilde \theta \leq 1$.  The results show that when $\eta=0.005$, for
more than $99.1\%$ of the rank distributions, $\tilde \theta$ is
larger than $0.96$; for all the rank distributions the smallest
$\tilde \theta$ is $0.9057$.  The figures in Fig.~\ref{fig:cdf}
clearly show the trend that when $\eta$ becomes smaller, $\hat\theta$
becomes larger for the same rank distribution.  Note that for these
rank distributions, the ratio $\sum_{r} r \hbar_r/\sum_r rh_r$ are all
larger than $0.999$, so the upper bound in Theorem~\ref{lemma:ub2} is
indeed very close to the capacity.

\subsection{Optimizations for Multiple Rank Distributions}
\label{sec:opt2}

In the previous part of this section, we consider how to find an
optimal degree distribution for a single rank distribution. For many
scenarios, however, we need a degree distribution that is good for
multiple rank distributions.  In a general multicast problem, the rank
distributions observed by the destination nodes can be different.
Even for a single destination node, the empirical rank distribution
may not always converge to the same value.  We discuss the degree
distributions for multiple rank distributions in
the remaining part of this section.

\begin{figure}
  \centering
  \begin{tikzpicture}
     \node[dot] (s) at(-2,0) {$s$};
     \node[dot] (a) at(0,0) {$a$} edge[<-] (s);
     \node[dot] (t) at(2,0) {$t_2$} edge[<-] (a);
     \node[dot] (t) at(2,-1) {$t_3$} edge[<-] (a);
     \node[dot] (t) at(2,1) {$t_1$} edge[<-] (a);
  \end{tikzpicture}
  \caption{In this network, node $s$ is the source node. Node $t_1,
    t_2$ and $t_3$ are the destination nodes. Node
    $a$ is the intermediate node that does not demand the file. All
    links are capable of transmitting one packet per use. The link
    $(s,a)$ has 
    packet loss rate $0.2$. The links $(s,t_i), i=1,2,3$ have packet
    loss rate $0.1$, $0.2$ and $0.3$, respectively.}
  \label{fig:oneonemany}
\end{figure}
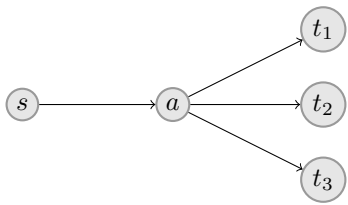

To illustrate the discussion, we extend the three-node network in
Fig.~\ref{fig:three} with two more destination nodes as shown in
Fig.~\ref{fig:oneonemany}. In this network, node $a$ transmits the
same packets on its three outgoing links, but these links have
different loss rates.  Fixing $M=16$, $q=256$ and certain inner code
in node $a$ (see the inner code to be defined in
Section~\ref{sec:line}), we obtain the rank distributions $h^i$ for
node $t_i$, $i=1,2,3$ in Table~\ref{table:rankd}.

\begin{table}
  \centering
  \caption{The rank distributions for the three destination nodes in Fig.~\ref{fig:oneonemany}.}
  \label{table:rankd}
  
  \begin{filecontents*}{data/rankdist3.txt}
   0.0000000e+00   6.5537000e-12   1.3107200e-11   4.3112257e-09
   1.0000000e+00   4.1944480e-10   8.3886080e-10   1.6112719e-07
   2.0000000e+00   1.2583884e-08   2.5165824e-08   2.8249687e-06
   3.0000000e+00   2.3492185e-07   4.6976203e-07   3.0858632e-05
   4.0000000e+00   3.0546474e-06   6.1068993e-06   2.3528289e-04
   5.0000000e+00   2.9338944e-05   5.8625437e-05   1.3297565e-03
   6.0000000e+00   2.1538861e-04   4.2987193e-04   5.7765384e-03
   7.0000000e+00   1.2338250e-03   2.4547213e-03   1.9737228e-02
   8.0000000e+00   5.5826374e-03   1.1009799e-02   5.3729648e-02
   9.0000000e+00   2.0084543e-02   3.8656509e-02   1.1649108e-01
   1.0000000e+01   5.7572891e-02   1.0414401e-01   1.9688036e-01
   1.1000000e+01   1.3064697e-01   2.0620263e-01   2.4681369e-01
   1.2000000e+01   2.2755184e-01   2.7951204e-01   2.1205998e-01
   1.3000000e+01   2.7961564e-01   2.3390669e-01   1.1201435e-01
   1.4000000e+01   2.0516803e-01   1.0386128e-01   3.1233438e-02
   1.5000000e+01   6.7100257e-02   1.8968058e-02   3.5716267e-03
   1.6000000e+01   5.1953344e-03   7.8917469e-04   9.3175556e-05
\end{filecontents*}
\pgfplotstabletypeset[
  every head row/.style={
    before row=\toprule,after row=\midrule,},
  every last row/.style={
    after row=\bottomrule},
  every even row/.style={
    before row={\rowcolor[gray]{0.9}}},
  columns/0/.style={
    column name=rank},
  columns/1/.style={
    column name=$h^1$,fixed,precision=4},
  columns/2/.style={
    column name=$h^2$,fixed,precision=4},
  columns/3/.style={
    column name=$h^3$,fixed,precision=4}
  ]{data/rankdist3.txt}
\end{table}

Let $\mathcal{H}$ be a set of rank distributions. Consider a degree
distribution $\Psi$ and 
$\theta_h, h\in \mathcal{H}$ satisfying the following set of constraints:
\begin{equation} \label{eq:cons29}
   \Omega(x;\Psi, \hbar(h)) + \theta_h \ln(1-x) \geq 0,\ \forall
   x \in [0, \bar\eta],\  \forall h \in \mathcal{H},
\end{equation}
where $\hbar(h)=(\hbar_i(h),i=1,\ldots,M)$.  Then for each rank
distribution $h\in\mathcal{H}$, rate $\bar\eta\theta_h$ is
$\bar\eta$-achievable by the BATS code with degree distribution
$\Psi$. 
For the above example, see the maximum $\bar\eta$-achievable rates
evaluated in Table~\ref{tab:rankdegree}.  The observation is that the
degree distribution optimized for one rank distribution may not have a
good performance for the other rank distributions: The degree
distributions optimized for destination node $t_1$ and $t_2$ have poor
performance for destination node $t_3$.

\begin{table}
  \centering
  \caption{The achievable rates for different pairs of rank
    distributions and degree
    distributions. For each rank distribution in the first row
    and each degree distribution in the first column, we evaluate the
    maximum $0.99$-achievable rate in the table. For $i=1,2,3$, $\Phi^i$ is obtained by solving \eqref{eq:op1} with
    $h^i$ in place of $h$. $\Psi^{3}$ can also be obtained by solving
    \eqref{eq:op2u} with $\{h^1,h^2,h^3\}$ in place of $\mathcal{H}$. 
    $\Psi^{\text{max-perc}}$ is obtain by solving \eqref{eq:op3} with $\{h^1,h^2,h^3\}$ in place of $\mathcal{H}$.}
  \label{tab:rankdegree}
  \begin{tabular}{c||c|c|c}
    \hline
    & $h^1$ & $h^2$ & $h^3$ \\
    \hline \hline
    $\sum_{i}i\hbar_i$ & 12.57 & 11.91 & 10.83 \\
    \hline
    $\Psi^1$ & 12.55 & 6.10 & 1.77 \\
    \hline
    $\Psi^2$ & 11.96 & 11.89 & 4.79 \\
    \hline
    $\Psi^3$ & 10.99 & 10.95 & 10.81 \\
    \hline
    $\Psi^{\text{max-perc}}$ & 11.94 & 11.35 & 10.28 \\
    \hline
  \end{tabular}
\end{table}

There are different criteria to optimize the degree distribution for
a set of rank distributions. Here we discuss two of them as examples.
One performance metric of interest is the multicast rate, which is a
rate that is achievable by all the rank distributions.  We can find
the maximum multicast rate for all the rank distributions in
$\mathcal{H}$ by solving the following optimization problem:
\begin{equation}
 \label{eq:op2u} \tag{P2}
 \max \theta \quad \text{s.t.} \left\lbrace
 \begin{IEEEeqnarraybox}[][c]{l}
   \Omega(x;\hbar(h),\Psi) + \theta \ln(1-x) \geq 0,\ \forall
   x \in [0, \bar\eta], \ \forall h \in \mathcal{H},\\
   \Psi_i \geq 0, \ \sum_i\Psi_i=1.
  \end{IEEEeqnarraybox}
  \right.
\end{equation}
Denote by $\hat\theta_{\mathcal{H}}$ the maxima of \eqref{eq:op2u}
w.r.t. $\mathcal{H}$.  By the upper bound discussed in
Section~\ref{sec:achieve}, $\bar\eta\hat\theta_{\mathcal{H}}$ should
be less than the minimum expected rank among all the rank
distributions in $\mathcal{H}$, denoted by $\bar h_{\mathcal{H}}$.
For the example that $\mathcal{H}=\{h^1,h^2,h^3\}$, the optimal degree
distribution of \eqref{eq:op2u} is exactly $\Psi^3$. Since nodes $t_1$
and $t_2$ can emulate the packet loss rate of node $t_3$, the
multicast rate of the BATS code is bounded by node $t_3$.  So in this
case, BATS codes can achieve a multicast rate very close to $\bar h_{\mathcal{H}}$.

\begin{figure}[tb]
  \centering
  \begin{tikzpicture}[scale=1]
    \begin{axis}[grid=both,
      xmin=0, xmax=16,
      ymin=0, ymax=16,
      xlabel=$\mu$, ylabel=Rate,
      ytick={0,2,4,...,16},
      legend pos=south east]
       \addplot+[mark=] table {data/example3.txt};
       \addlegendentry{$\bar\eta\hat{\theta}_{\mathcal{B}(\mu)}$}
       \addplot+[dashed,mark=] coordinates {(0,0) (16,16)};
       \addlegendentry{$\mu$}
    \end{axis}
  \end{tikzpicture}
  \caption{The optimal values of \eqref{eq:op2u} w.r.t.\ 
    $\mathcal{B}(\mu)$, where
    $M=16$, $q=2^8$ and $\bar\eta=0.99$.}
  \label{fig:op2u}
\end{figure}
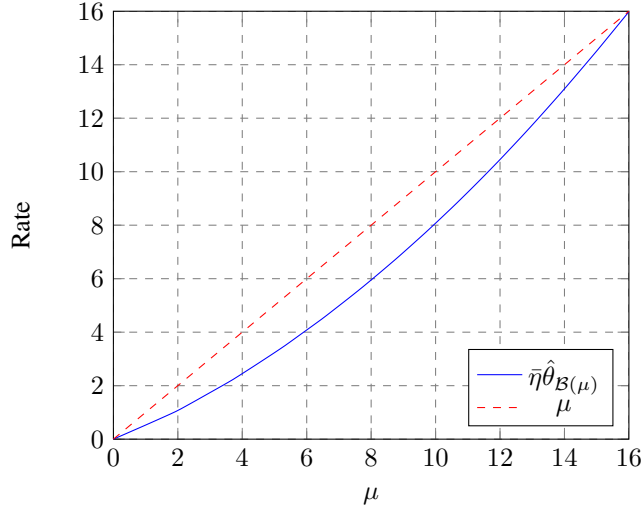

In general, however, $\bar\eta\hat\theta_{\mathcal{H}}$ may not be very close to $\bar h_{\mathcal{H}}$.  The maximum gap between $\bar\eta\hat\theta_{\mathcal{H}}$ and $\bar h_{\mathcal{H}}$ can be
obtained numerically.  For any real value $\mu$, $0\leq \mu \leq M$, define
  \begin{equation*}
    \mathcal{B}(\mu)  =  \left\{ (h_0,h_1,\ldots,h_M) :\sum_{i=1}^M ih_i \geq \mu, \sum_{i=1}^Mh_i=1, h_i\geq 0, i=0,\ldots,M \right\}.
  \end{equation*}
  The set $\mathcal{B}(\mu)$ includes all the rank distributions that
  can potentially support rate $\mu$.  Since using more rank
  distributions can only give smaller optimal values, solving
  \eqref{eq:op2u} w.r.t.\ $\mathcal{B}(\mu)$ gives us a guaranteed
  multicast rate that is achievable by BATS codes with BP decoding for
  any set of rank distributions $\mathcal{H}$ with $\bar
  h_{\mathcal{H}} = \mu$.  Directly solving \eqref{eq:op2u} w.r.t.\
  $\mathcal{B}(\mu)$ is difficult since $\mathcal{B}(\mu)$ includes
  infinitely many of rank distributions. Using the techniques developed
  in \cite{yang11ddo}, the set $\mathcal{B}(\mu)$ can be reduced to a
  finite set, and hence \eqref{eq:op2u} can be solved efficiently.
  See Fig.~\ref{fig:op2u} for
  $\bar\eta\hat{\theta}_{\mathcal{B}(\mu)}$ when $M=16$, $q=256$ and
  $\bar\eta=0.99$.  For example,
  $\bar\eta\hat{\theta}_{\mathcal{B}(10)} = 8.10$.

The degree distribution obtained using \eqref{eq:op2u} may not be fair
for all the destination nodes. For the degree distribution optimized
using \eqref{eq:op2u}, nodes $t_1$ and $t_2$ do not achieve a rate
much higher than node $t_2$ though they have much lower loss rate than
node $t_2$ (ref. Table~\ref{tab:rankdegree}).  To resolve this issue,
we can find the percentage of $\sum_{i}i\hbar_i(h)$ that is achievable
for all the rank distributions $h$ in $\mathcal{H}$ using the
following optimization:
\begin{equation}
 \label{eq:op3} \tag{P3}
 \max \alpha \quad \text{s.t.} \left\lbrace
 \begin{IEEEeqnarraybox}[][c]{l}
   \Omega(x;\hbar(h),\Psi) + \alpha \sum_{i}i\hbar_i(h) \ln(1-x) \geq
   0,\ \forall
   x \in [0, \bar\eta], \ \forall h \in \mathcal{H},\\
   \Psi_i \geq 0, \ \sum_i\Psi_i=1.
  \end{IEEEeqnarraybox}
  \right.
\end{equation}
Denote by $\hat{\alpha}$ the maxima of \eqref{eq:op3}.
When $\mathcal{H}=\{h^1,h^2,h^3\}$ and $\bar\eta = 0.99$, the
percentage is 94.9 (the optimal value of \eqref{eq:op3} multiplied by
$\bar\eta$).  The performance of the optimal degree distribution of
\eqref{eq:op3} is shown in the last row of Table~\ref{tab:rankdegree}.
BATS codes with this degree distribution achieves 95.0, 95.3 and 94.9
percentage of $\sum_{i}i \hbar_i$ for sink nodes $t_1$, $t_2$ and
$t_3$, respectively.

In general, BATS codes are not universal. There does not exist a rank
distribution that can achieve rates close to $\sum_{i}i\hbar(h)$ for
all rank distributions for a given batch size $M$, except for $M=1$,
the case of LT/Raptor codes. In Table~\ref{tab:maxratio}, we give the
optimal values of \eqref{eq:op3} (multiplied by $\bar\eta$) with
$\mathcal{H}$ being the set of all the rank distributions for batch
size $1,2,4,\ldots,64$.  Take $M=16$ as an
example. The value $\bar \eta \hat \alpha = 0.5274$ implies a worst
guaranteed rateless rate for an arbitrary number of destination nodes with
arbitrary empirical rank distributions: A destination node can decode
the original file with high probability after receiving $n$ batches such that $0.5274
\sum_{i=1}^n \rank(\bH_i)$ is larger than the number of original input
packets, where $\bH_i$ is the transfer matrix
of the $i$th batch. When
the possible empirical rank distributions are in a smaller set, the
optimal value of \eqref{eq:op3} can be much larger, as in the network
with three destination nodes.

\begin{table}
  \centering
  \caption{The maximum value $\hat\alpha$ of \eqref{eq:op3} when $\mathcal{H}$ is the set of all rank distributions
  for a given batch size. Here $\bar\eta = 0.99$}
  \label{tab:maxratio}
  \begin{tabular}{c||c|c|c|c|c|c|c}
    \hline
    $M$ & 1 & 2 & 4 & 8 & 16 & 32 & 64 \\
    \hline
    $\bar\eta \hat\alpha$ & 0.9942 & 0.8383 & 0.7068 & 0.6060 & 0.5274
    & 0.4657 & 0.4165 \\
    \hline
  \end{tabular}
\end{table}

Using different objective functions and constraints, other
optimization problems can be formulated to optimize a degree
distribution for a set of rank distributions. For
example, we can optimize the average rate and average completion time
of all the destination nodes. Readers are referred to \cite{yang11ddo}
for more degree distribution optimization problems and the techniques
to solve these problems.

\section{Practical Batch Encoding and Decoding Designs}
\label{sec:finite}

Asymptotic performance of BATS codes has been studied in the previous
sections. Now we look at BATS code with finite block
lengths. We highlight the issues of BATS codes with finite block
lengths and discuss the techniques that can resolve the issues. 

\subsection{Overhead and Rate}

We give an alternative and convenient way to evaluate the performance
of a BATS code with finite block lengths.  We define two kinds of
overheads, which is related to the outer code and inner code
respectively, and discuss their relationship with coding rate.

Suppose that a destination node decodes successfully after receiving
$n$ batches with transfer matrices $\{\bH_i, i=1,\ldots, n\}$. If the
coding vector of a received packet is linearly dependent with those of
the other received packets of the same batch, this packet is redundant
and can be discarded by the decoder.  Therefore we define the
\emph{receiving overhead} as $\text{RO} = \sum_{i=1}^n
[\text{col}(\bH_i) - \rank(\bH_i)]$, where $\text{col}(\bH)$ is the
number of columns of $\bH$. The receiving overhead is generated inside
the networks by the inner code, and hence cannot be reduced
by the design of batch encoding and decoding. We should design the
inner code to reduce the receiving overhead, but it may not
always be necessary to reduce the receiving overhead to a value close
to zero (see example designs in the next section).

Define the \emph{coding overhead} as $\text{CO} = \sum_{i=1}^n
\rank(\bH_i) - K'$, where $K'$ is the number of original input packets
before precoding. We should design batch encoding and decoding schemes
such that the coding overhead is as small as possible. The
\emph{coding rate observed by the destination node} is $\text{CR} =
\frac{K'}{\sum_{i=1}^n \text{col}(\bH_i)} =
\frac{K'}{\text{RO}+\text{CO}+K'}$. For an optimal code, we have
$\text{CO}\rightarrow 0$ and hence $\text{CR} \rightarrow
\frac{\sum_{i=1}^n \rank(\bH_i)}{\sum_{i=1}^n\text{col}(\bH_i)}$, the
normalized average rank of the transfer matrices.

\subsection{Inactivation Decoding}

BP decoding is an efficient way to decode BATS code, but it may not
always achieve the maximum rates, for example, in case of multiple
destination nodes with different rank distributions. Even for a single
destination node, BP decoding stops with high probability when both
the number of input packets and the coding overhead are relatively small.  Instead
of tolerating large coding overheads, we can continue the decoding
using Gaussian elimination after BP decoding stops. But Gaussian
elimination has a much higher complexity. A better way is
\emph{inactivation decoding}, which is an efficient way to solve
sparse linear systems \cite{Lamacchia90, Pomerance92}.  Inactivation
decoding has been used for LT/Raptor codes \cite{inactivation,
  Raptormono}, and similar algorithm has been used for efficient
encoding of LDPC codes \cite{Richardson01}.

Recall that BP decoding stops when there are no decodable batches.  In
inactivation decoding, when there are no decodable batches at time
$t$, we instead pick an undecoded input packet $b_k$ and mark it as
\emph{inactive}. We substitute the inactive packet $b_k$ into the
batches like a decoded packet, except that $b_k$ is an indeterminate.
For example, if $b_k$ is a contributor of batch $i$, we express
the components of $\mathbf{Y}_i^{t+1} = \mathbf{Y}_i^t - b_k g$ as
polynomials in $b_k$.  The decoding process is repeated until
all input packets are either decoded or inactive.  The inactive input
packets can be recovered by solving a linear system of equations using
Gaussian elimination.  %
In a nutshell, inactivation decoding trades computation cost (decoding inactive input
symbols using Gaussian elimination) with coding overhead.

\subsection{Design of Degree Distributions}
 
The degree distribution obtained using the optimization \eqref{eq:op1}
is guaranteed to have good performance for sufficiently large block
length by Theorem~\ref{the:1}, but 
performs poorly for relative small block length, e.g., from several
hundreds to several thousands input packets. We can use the heuristical
method introduced in \cite{ng13ff} to design the degree distribution
of a BATS code with finite block length:
\begin{equation}
\max\theta \ \text{s.t.} \left\lbrace
\begin{IEEEeqnarraybox}[][c]{l}
\Omega(x)+\theta\left[ \ln(1-x) - \frac{c}{K} (1-x)^{c'} \right] \geq 0\ \text{ for } 0 \leq x \leq 1-\eta\\
\sum_d \Psi_d = 1, \ \Psi_d \geq 0  \text{ for } d=1, \ldots, D,
\end{IEEEeqnarraybox}
\right. \label{eq:op4} \tag{P4}
\end{equation}
where $c$ and $c'$ are parameters that we can tune. The intuition
behind the above optimization of degree distribution can be found in
\cite[Section VII]{ng13ff}.

For given values of $c$ and $c'$, optimization \eqref{eq:op4} provides
us with candidate degree distributions that could have better
performance. The important part is how to evaluate a degree
distribution. This task can be done using simulations, but it can be
done more efficiently with the iterative formulae developed in
\cite{ng13,ng13ff}.  For any given values of $K$, $n$, degree
distribution and rank distribution, these formulae
calculate the exact error probability for BP decoding and expected
number of inactive packets for inactivation decoding. 
Readers are referred to \cite{ng13ff} for detailed discussions.

\subsection{Precode Design}
\label{sec:precodedesign}

Due to similar requirements, the precode for Raptor codes in
\cite{Raptormono, raptorq} can be applied to BATS without much
modifications. Major techniques include high-density parity check and
permanent inactivation, which help to reduce the coding overhead of
inactivation decoding. Readers can
find the detailed discussion of these techniques in \cite{Raptormono}.

\bigskip

As a summary, using the above techniques, it is possible to design
BATS codes with very low coding overhead for finite block lengths when
the degree distribution is optimized for a rank distribution. We
use the length-$4$ line network (to be introduced formally in the next
section) as an example to evaluate the finite length performance of a
BATS code with inactivation decoding. The results are in
Table~\ref{tab:1}, where the average overhead is less than 3 packets per 1600 packets.

\begin{table}
  \centering
  \caption{Numerical results of BATS codes for the length-$4$ line network. We set
    $M=32$ and $q=256$.}
  \label{tab:1}
  \begin{tabular}{c|c|c|c|c|c|c|c|c|c}
    \hline
    \multirow{2}{*}{$K$} & \multicolumn{3}{c|}{coding overhead} &
    \multicolumn{3}{c|}{inactivation no.} & \multicolumn{3}{c}{receiving overhead}\\
    \cline{2-10}
    & average & max & min & average & max  & min & average & max  & min\\
    \hline \hline
    1600 & 2.04 & 16 & 0 & 94.0 & 119 & 72 & 599.5 & 673 & 532 \\
    \hline
    8000 & 6.30 & 77 & 0 & 215.5 & 268 & 179 & 3015.7 & 3183 & 2865 \\
    \hline
    16000 & 26.58 & 1089 & 0 & 352.2 & 379 & 302 & 6041.6 & 6469 & 5788 \\
    \hline
  \end{tabular}
\end{table}

\section{Examples of BATS Code Applications}
\label{sec:example}

In this section, we use several examples to illustrate how to apply
BATS codes in erasure networks, where each (network) link can transmit
\emph{one} packet in a time slot subject to a certain packet erasure
probability.  If not erased, the packet will be correctly received.
We say a network has \emph{homogeneous links} if all network links
have the same erasure probability, and has \emph{heterogeneous links}
otherwise.  Unless otherwise specified, network link transmission is
instantaneous.  We will focus on how to design the inner code
including cache management and batch scheduling at the intermediate
nodes.

\subsection{Line Networks}
\label{sec:line}

A line network of length $k$ is formed by a sequence of $k+1$ nodes
$\{v_0,v_1,\ldots,v_k\}$, where the first node $v_0$ is the source
node and the last node $v_k$ is the destination node. There are only
network links between two consecutive nodes. The network in
Fig.~\ref{fig:three} is a line network of length $2$.  We first study
line networks with \emph{homogeneous links} and then extend the
results to general line networks. Suppose that all the links in the
line network has the same link erasure probability $\epsilon$.  When
there is no computation and storage constraints at the intermediate
network nodes, the min-cut capacity of the line network with length
$k$ is $1-\epsilon$ packet per use for any $k>0$. Here one use of the
network means the use of each network link at most once; it is
possible to transmit nothing on a network link in a particular time
slot.  We apply the following BATS code scheme for line networks.

\begin{scheme}[Line network] \label{sch:1}
In this BATS code scheme for a line network, 
the source
node generates batches and transmits a packet in each time slot. The
$M$ packets of a batch is transmitted in $M$ consecutive time slots,
and the batches are transmitted according to the order in which they are
generated. The source node keeps transmitting batches until the
destination node decodes successfully. No feedback is required except for
the notification of successful decoding from the destination node.

In the first $M$ time slots, node $v_1$ can potentially receive $M$
packets of the first batch. In the first $M-1$ time slots, node $v_1$
saves the received packets in its buffer but transmits nothing. In the
$M$th time slot, node $v_1$ generates $M$ coded packets using random
linear coding on the packets in its buffer and the packet just
received, if any, which are all in the same batch.  After generating
the $M$ coded packets, the original received packets in the buffer are
deleted. Node $v_1$ then transmits one of the coded packets and saves
the remaining $M-1$ coded packets in its buffer.  In each of the
following $M-1$ time slots, node $v_1$ transmits one of the remaining
coded packets of the first batch and then deletes in the buffer the transmitted
packet immediately. During these time slots, if node
$v_1$ receives a new packet (of the $2$nd batch), the new packet is
saved in the buffer. At the $2M$th time slot, node $v_1$ repeats its
operations on the first batch at the $M$th time slot, so on and so
forth.  All the other intermediate nodes apply the same operations as
node $v_1$.
\end{scheme}

In the above scheme, each intermediate node caches at most $M-1$
packets in the buffer.  There is a delay for each intermediate node:
node $v_i$ can only start to receive packets after $(i-1)(M-1)$ time
slots. For a network of fixed length, the delay is neglectable
compared with the total transmission time when the file size is large.
The buffer at the intermediate nodes may be better managed to improve
the rate and/or to reduce the delay, but the scheme we defined here is
easy to analyze and is asymptotically optimal.

The transfer matrices of all batches are i.i.d. and can be expressed
explicitly. The transmission of a batch through a network link can be
modelled by an $M\times M$ random \emph{diagonal} matrix $E$ with
independent components, where a diagonal component is $0$ with
probability $\epsilon$ and is $1$ with probability $1-\epsilon$. The
network coding at an intermediate node for a batch is given by a
totally random $M\times M$ matrix $\Phi$. The transfer matrix
$H^{(1)}$ for the unit-length line network is $H^{(1)} = E_1$, where
$E_1$ has the same distribution as $E$.  For $k>1$, the transfer
matrix $H^{(k)}$ for the $k$-length line network can be expressed as
\begin{equation*}
  H^{(k)} = H^{(k-1)} \Phi_{k-1} E_k,
\end{equation*}
where $\Phi_{k-1}$ has the same distribution as $\Phi$ and $E_k$ has
the same distribution as $E$. Further, $\Phi_{1}, \ldots, \Phi_{k-1}$,
$E_1, \ldots, E_k$ are mutually independent.

The rank distribution of the transfer matrix $H^{(k)}$ can be calculated
recursively. Let $h^{(k)} = (h^{(k)}_0,\ldots,h^{(k)}_M)$ be the rank
distribution of $H^{(k)}$. First
\begin{equation*}
  h^{(1)}_r = \binom{M}{r} (1-\epsilon)^r \epsilon^{M-r}, \quad r=0,1,\ldots,M.
\end{equation*}
Using \eqref{eq:cp1} we obtain that for $k>1$,
\begin{equation*}
  h^{(k)}_r = \sum_{i=r}^M \sum_{j=r}^M  h^{(k-1)}_i \binom{M}{j} (1-\epsilon)^j \epsilon^{M-j} \cmatt{i,j}{r} , \quad r=0,1,\ldots,M.
\end{equation*}
When $M=1$, the BATS code scheme for line networks degenerates to an
LT/Raptor code scheme with forwarding at the intermediate nodes. The
achievable rate for the length-$k$ line network is $(1-\epsilon)^k$, i.e., the
rate decreases exponentially fast with the network length.  

We know from the previous discussion that the normalized expected rank
$\sum_{r}r h^{(k)}_r/M$ can be approached by BATS codes.  When
$M$ tends to infinity, the normalized expected rank will converge to
$1-\epsilon$, which can be verified using the recursive formula of the
rank distributions. Therefore, for line networks with link erasure
probability $\epsilon$, Scheme~\ref{sch:1} can achieve a normalized
rate very close to $1-\epsilon$ when $M$ is sufficiently large. If the
coding vector is included in the packets, $T$ should also be large
so that $M/T$ is small.  However, we are more
interested in the performance for small values of $M$, which can be
characterized numerically.

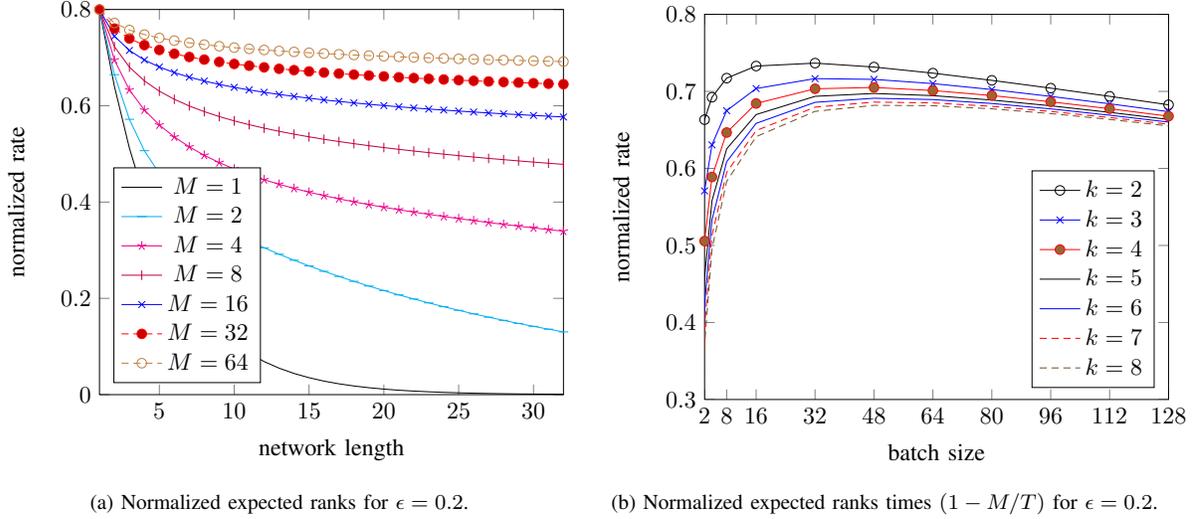
\begin{figure}[tb]
  \centering
  \begin{subfigure}[b]{0.45\textwidth}
    \centering
    \begin{tikzpicture}[scale=0.9]
    \begin{axis}[
      xmin=1, xmax=32,
      ymin=0, ymax=0.8,
      xlabel=network length, ylabel=normalized rate,
      legend pos=south west]
       \addplot+[color=black,no marks] table[x=idx,y=1]
       {data/rankdistline.txt};
       \addlegendentry{$M=1$}
       \addplot+[color=cyan,mark=-] table[x=idx,y=2] {data/rankdistline.txt};
       \addlegendentry{$M=2$}
       \addplot+[color=magenta,mark=star] table[x=idx,y=4] {data/rankdistline.txt};
       \addlegendentry{$M=4$}
       \addplot+[color=purple,mark=|] table[x=idx,y=8] {data/rankdistline.txt};
       \addlegendentry{$M=8$}
       \addplot+[color=blue,mark=x] table[x=idx,y=16] {data/rankdistline.txt};
       \addlegendentry{$M=16$}
       \addplot+[color=red,mark=*] table[x=idx,y=32] {data/rankdistline.txt};
       \addlegendentry{$M=32$}
       \addplot+[color=brown,mark=o] table[x=idx,y=64] {data/rankdistline.txt};
       \addlegendentry{$M=64$}
    \end{axis}
  \end{tikzpicture}
  \caption{Normalized expected ranks for $\epsilon =
      0.2$.}
  \label{fig:linenetwork}
  \end{subfigure}
~~~
  \begin{subfigure}[b]{0.45\textwidth}
    \centering
    \begin{tikzpicture}[scale=0.9]
    \begin{axis}[
      xmin=2, xmax=128,
      xtick={2,8,16,32,...,128},
      ymin=0.3, ymax=0.8,
      xlabel=batch size, ylabel=normalized rate,
      legend pos=south east]
       \addplot+[color=black,mark=o] table[x=M,y=2]
       {data/linenetrate.txt};
       \addlegendentry{$k=2$}
       \addplot+[color=blue,mark=x] table[x=M,y=3] {data/linenetrate.txt};
       \addlegendentry{$k=3$}
       \addplot+[color=red,mark=*] table[x=M,y=4] {data/linenetrate.txt};
       \addlegendentry{$k=4$}
       \addplot+[no marks] table[x=M,y=5] {data/linenetrate.txt};
       \addlegendentry{$k=5$}
       \addplot+[no marks] table[x=M,y=6] {data/linenetrate.txt};
       \addlegendentry{$k=6$}
       \addplot+[no marks] table[x=M,y=7] {data/linenetrate.txt};
       \addlegendentry{$k=7$}
       \addplot+[no marks] table[x=M,y=8] {data/linenetrate.txt};
       \addlegendentry{$k=8$}
    \end{axis}
  \end{tikzpicture}
    \caption{Normalized expected ranks times $(1-M/T)$ for $\epsilon =
      0.2$.}
  \label{fig:linenetwork2}
  \end{subfigure}
  \caption{Numerical results for line networks. The field size $q$ is
    $2^8$.}
\end{figure}

We calculate the normalized expected rank $\sum_{r}r h^{(k)}_r/M$ for
$\epsilon=0.2$ and field size $q=256$ in
Fig.~\ref{fig:linenetwork}. Compared with $M=1$, the normalized
expected rank decreases slowly as the network length increases when
$M\geq 2$.  For a fixed network length, Fig.~\ref{fig:linenetwork}
also illustrates the tradeoff between the batch size and the maximum
achievable rates of BATS codes (without considering the coding vector
overhead, or assuming $T$ is much larger than $M$). We see that when
$M$ is larger than $32$, using a larger batch size only gives a
marginal rate gain (but increases significantly the computation cost).

The gain by using a larger $M$ can be offset by the coding vector
overhead.  If we include the coding vector overhead, the normalized
rate of BATS codes can be very close to $(1-M/T) \sum_{r}r
h^{(k)}_r/M$. In the extreme case that $M=T$, the achievable rate
becomes zero. We calculate the value of $(1-M/T) \sum_{r}r
h^{(k)}_r/M$ for $T=1024$, $\epsilon=0.2$ and $q=256$ in
Fig.~\ref{fig:linenetwork2}. These values of $T$ and $q$ correspond to
a packets size of 1 KB.  It can be seen from the plot that when
$\epsilon=0.2$, a small batch size roughly equal to $32$ is almost
rate-optimal for practical parameters.

If $\epsilon$ is large, e.g., $0.9$, however, a much larger batch
size, e.g., $200$, will be required so that the normalized expected
rank approaches $1-\epsilon$. But a large batch size results in a
large coding vector overhead.  We introduce a technique such that
small batch sizes can still be used for high erasure
probabilities. Suppose that $M$ is a large enough batch size such that
the normalized expected rank is close $1-\epsilon$.  Let $\tilde M =
M(1-\epsilon+\delta)$ for certain small positive value $\delta$. For
example, $\tilde M = 30$ when $\epsilon = 0.9$, $M=200$ and
$\delta=0.05$. We modify Scheme~\ref{sch:1} by using an outer code
with batch size $\tilde M$ to replace the outer code with batch size
$M$: For each batch $X$ of size $\tilde M$, the source node generates
$M$ packets for transmission by multiplying $X$ with an $\tilde
M\times M$ totally random matrix $\Psi_{\tilde M\times M}$. The inner
code does not change: the batch size is still $M$.  The destination node
decodes by using batch size $\tilde M$.

The effectiveness of the above technique is explained as follows.
Let $H$ be the transfer matrix of a batch for the outer code with
batch size $M$. In the above modified Scheme~\ref{sch:1}, the transfer
matrix of a batch for the outer code with batch size $\tilde M$ can be
expressed as $\Psi_{\tilde M\times M}H$. We know that in
Scheme~\ref{sch:1}, the rank of $H$ is smaller than $\tilde M =
M(1-\epsilon+\delta)$ with high probability when $M$ is large. Thus,
the expected ranks of $H$ and $\Psi_{\tilde M\times M}H$ converges to
the same value in probability as $M$ tends to infinity. Therefore, the
asymptotic performance of the outer codes is not sacrificed by using a
small batch size. 

The above technique can in principle be applied for all values of
$\epsilon$. However, for small values of $\epsilon$, the advantage of
using the technique is small.

Scheme~\ref{sch:1} does not depend on the erasure probability of
network links, so it can also be applied to a line network with
heterogeneous links. Consider a length-$k$ line network where the
maximum link erasure probability among all links is $\epsilon$. The
min-cut capacity of this network is $1-\epsilon$. The expected rank of
the transfer matrix of this network is less than the one of the
length-$k$ line netwrok where the erasure probabilities of all links
are $\epsilon$. Therefore, the normalized expected rank will converge
to $1-\epsilon$ when $M$ tends to infinity, and hence
Scheme~\ref{sch:1} can achieve a normalized rate very close to
$1-\epsilon$ when $M$ is sufficiently large for a line network.

\subsection{Unicast Networks}

A unicast network is represent by a directed acyclic graph with one
source node and one destination node. We first provide two BATS code
schemes for unicast networks with \emph{homogeneous links}, and then
discuss how to extend the schemes to unicast networks with heterogeneous
links.  One way to apply Scheme~\ref{sch:1} to unicast
networks with homogeneous links is as follows.

\begin{scheme}[Unicast]
  \label{sch:2}
  Consider a unicast network with homogeneous links.  Find 
  $L$ edge-disjoint paths from the source node to the destination
  node, and separate the input packets into $L$ groups, each of which
  is associated with a path. The source node encodes each group of the
  input packets using a BATS code, and transmits all the batches on
  the associated path. An intermediate node on that path runs an instance of
  the inner code of the BATS code defined in Scheme~\ref{sch:1}.  The
  destination node decodes the packets received from a path to
  recover a group of input packets.
\end{scheme}

For a unicast network with link erasure probability $\epsilon$ for all
links, the min-cut capacity is $(1-\epsilon)L^*$, where $L^*$ is the
maximum number of edge-disjoint paths from the source node to the
destination node.  Since Scheme~\ref{sch:2} is equivalent to applying
Scheme~\ref{sch:1} on multiple line networks, the normalized expected
rank of the transfer matrix for each path converges to $1-\epsilon$
as $M$ tends to infinity. Therefore, Scheme~\ref{sch:2} can achieve a
rate very close to the min-cut capacity by optimizing the degree
distribution for each path separately. However, a better scheme can be
obtained by encoding and decoding the batches for different paths
jointly.

\begin{scheme}[Unicast]
  \label{sch:3}
  This scheme for a unicast network with homogeneous links is same as
  Scheme~\ref{sch:2} except that the source node encodes all the input
  packets using a BATS code. The batches are grouped into sets of $L$
  sequentially. Each set is transmitted on the $L$ paths in $M$ time
  slots, with each batch in the set transmitted on a distinct path.
\end{scheme}

The rank distribution for the above scheme is the rank distributions
averaged over all the paths. So, the normalized expected rank of the
transfer matrix converges to $1-\epsilon$ as $M$ tends
to infinity.  Now we consider a general unicast network with
heterogeneous links.  We apply the above BATS code scheme to the
 unicast network in three steps:
\begin{enumerate}
\item Obtain a unicast network $G^*$ with homogeneous links that has the same
  min-cut as the original unicast network $G$. 
\item Apply Scheme~\ref{sch:3} on network $G^*$.
\item Convert the scheme on $G^*$ to one that can be used in network $G$ while preserving the performance.
\end{enumerate}
The second step is straightforward. The first and third
steps are explained as follows.

In the first step, assume that the link erasure probability are all
rational. Fix an integer $N$ such that $(1-\epsilon)N$ is an integer
for any erasure probability $\epsilon$ in a link of the
network. Network $G^*$ has the same set of nodes as network $G$. For
any link between nodes $a$ and $b$ in $G$ with erasure probability
$\epsilon$, we have a set of $(1-\epsilon)N$ parallel links between
nodes $a$ and $b$ in $G^*$ with erasure probability $1-1/N$.  We call
network $G^*$ the \emph{homogenized network} of network $G$.  We can
check that the min-cut capacity of network $G$ and $G^*$ are the same.
Use the three-node network in Fig.~\ref{fig:three} as example.
Suppose that the two links $(s,a)$ and $(a,t)$ have erasure
probabilities $0.2$ and $0.1$, respectively. Let $N=10$. The
homogenized network of the three-node network has 8 parallel links
from node $s$ to node $a$ and 9 parallel links from node $a$ to node
$t$, where all the links have an erasure probability $0.9$.

In the third step, we convert Scheme~\ref{sch:3} on network $G^*$ to
one that can be used in the original network by emulating virtual
links in the network nodes in $G$. In the three-node network example, 
for link $(s,a)$,
node $s$ emulates $8$ virtual outgoing links and
node $a$ emulates $8$ virtual incoming links, each of which
corresponds to a virtual outgoing link of node $s$; for link $(a,t)$,
node $a$ emulates $9$ virtual outgoing links and
node $t$ emulates $9$ virtual incoming links, each of which
corresponds to a virtual outgoing link of node $a$. 
In each time slot, nodes $s$ and $a$ randomly
choose one of their virtual outgoing links, transmit the packet on
that virtual link on the original outgoing link, and delete the packets on
the other virtual outgoing links.  We assume that the choice of virtual
outgoing links in node $s$ is known by node $a$ so that the received
packet of node $a$ from link $(s,a)$ can be associated with the corresponding
virtual incoming link. The same is assumed for nodes $a$ and $t$.
In a general network topology, a network node needs to maintain a set
of virtual outgoing (incoming) links for each original outgoing
(incoming) link.

The rank distribution induced by Scheme~\ref{sch:3} on network $G^*$
is the same as the modified scheme on the original network $G$. Therefore,
the BATS code scheme can achieve a rate very close to the min-cut
capacity of a unicast network when $M$ is sufficiently large. 

\subsection{Two-way Relay Networks}

The BATS code scheme for line networks can also be used in two-way
(multi-way) relay networks combined with existing schemes for the
relay node \cite{Katti2006, Zhang06}. A two-way relay network has
the same topology as the three-node network in Fig.~\ref{fig:three},
except that the two links share a common wireless channel and are
bidirectional. A packet transmitted by node $a$ can be received by
both nodes $s$ and $t$.  Nodes $s$ and $t$ each have a file to
transmit to each other. Since there is no direct link between nodes
$s$ and $t$, the transmission must go through node $a$. We introduce
two schemes for two-way relay networks, which may not have
homogeneous links.

\begin{scheme}[Two-way relay]
  \label{sch:4}
  Since each direction of the transmission in a two-way relay
  network is a length-2 line network, we use Scheme~\ref{sch:1}
  for each direction of the transmission. Node $a$
  runs two instances of the inner code defined in
  Scheme~\ref{sch:1}, each for one direction of transmission.  The
  channel is shared by the three nodes as follows: Nodes $s$ and $t$
  each transmit $M$ time slots and then node $a$ transmits $M$ time
  slots.  To get the benefit of network coding for the two-way relay
  network, an extra operation is required at node $a$: node $c$
  combines the batches from the two source nodes for transmission. At
  each of the time slots $M, 2M, \ldots$, node $a$ generates $M$
  recoded packets for a batch from node $s$ and $M$ recoded packets
  for a batch from node $t$. To transmit these $2M$ packets in $M$
  uses of the channel, node $a$ combines one packet from node $s$ and
  one packet from node $t$.  Nodes $s$ and $t$ can first cancel their
  own packets from the received packets and then run the BP decoding
  of the respective BATS codes.
\end{scheme}

Physical-layer network coding (PNC) \cite{Zhang06,popovski06} (also an
instance of compute-and-forward \cite{Nazer11}) can further reduce the
number of channel uses. The idea is to allow nodes $s$ and $t$ to
transmit simultaneously so that node $a$ receives a physical
superposition of the signals transmitted by nodes $s$ and $t$. Instead
of decoding the original packets transmitted, node $a$ tries to decode
a linear combination of these two packets. The decoding results will
be in one of the following cases: i) a linear combination is decoded
with nonzero coefficients for both packets (no erasure on both links);
ii) the packet of node $s$ or $t$ is decoded (an erasure on one of the
link); iii) nothing is decoded (erasures on both links).

\begin{scheme}[Two-way relay]
  \label{sch:5}
  In this scheme for a two-way relay network with physical-layer network
  coding, we schedule the nodes in a way such that nodes $s$ and $t$
  simultaneously transmit $M$ times slots and then node $a$ transmits
  $M$ time slots.  Node $a$ applies the inner code defined in
  Scheme~\ref{sch:1} by treating the packets decoded by PNC (cases i)
  and ii) above) during times slots $iM+1,\ldots iM+M$ as a batch,
  where $i=0, 1,\ldots$.  The operations in nodes $s$ and $t$ remain
  the same as Scheme~\ref{sch:4}.
\end{scheme}

\subsection{Tree Networks and CDNs}
\label{sec:tree}

We continue to discuss tree networks where the root is the source node
and all the leaves are destination nodes. We first consider a tree
network with \emph{homogeneous links}. Scheme~\ref{sch:1} can be
extended to such a tree network by allowing a node to transmit the
same packets on all its outgoing links. For each destination node, the
normalized expected rank of the transfer matrix induced by the inner
code of this scheme converges to the min-cut as the batch size tends
to infinity. A general performance metric of the BATS code scheme with
BP decoding is a tuple of achievable rates for all destination nodes,
which can be characterized numerically using the techniques introduced
in Section~\ref{sec:opt2}.  Since the rank distributions of different
destination nodes can be different, we need the degree distribution
optimization techniques introduced in Section~\ref{sec:opt2}.

This BATS code scheme with BP decoding can asymptotically achieve a
multicast rate very close to the multicast capacity of a tree network
with homogeneous links: When all the leaves have the same depth, the
rank distributions of all the destination nodes are the same, and
hence the degree distribution obtained by solving \eqref{eq:op1}
w.r.t. the common rank distribution can achieve a rate very close to
the common expected rank for all destination nodes.  When the leaves
of the tree network have different depth, the multicast capacity is
the min-cut capacity of the destination node $t$ with the largest
depth. Since a destination node with a smaller depth can emulate node
$t$, we can use the BATS code optimized for node $t$ to approach the
multicast capacity.

We now extend the tree network scheme to multicast networks with
homogeneous links,
which are represented by directed acyclic graphs with one source node and
multiple destination nodes. 

\begin{figure}
  \centering
  \begin{tikzpicture}[scale=1.8] 
	\node[dot] (s) at (0,0) {$s$}; 
	\node[dot] (a) at (-1,-0.8) {} edge [<-] (s); 
	\node[dot] (b) at (1,-0.8) {} edge [<-] (s); 
	\node[dot] (c) at (0,-0.8) {} edge [<-] (s);
	\node[dot] (d) at (0,-1.6) {} edge [<-] (a) edge [<-] (b); 
	\node[dot] (t) at (-1,-1.6) {} edge [<-] (c) edge [<-] (a); 
	\node[dot] (u) at (1,-1.6) {} edge [<-] (c) edge [<-] (b); 
	\end{tikzpicture} 
  \caption{Three-Layer network. Node $s$ in the top layer is the source node. Nodes in
    the middle layer are the intermediate nodes. Nodes in the bottom
    layer are the destination nodes.}
  \label{fig:threelayer}
\end{figure}
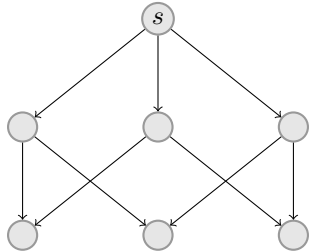

\begin{scheme}[Multicast]\label{sch:6}
  Consider a multicast network with homogeneous links.
  Find a set of edge-disjoint trees each of which has the source node
  as the root and a \emph{subset} of the destination nodes as the
  leaves. (The subsets of the destination nodes of different trees may
  be different and overlap.) Apply Scheme~\ref{sch:3} for unicast networks
  with the set of trees in place of the set of paths. For
  each tree, a node transmits the same packets to all its
  children. (See example below for the network in Fig.~\ref{fig:threelayer}.)
\end{scheme}

We can apply the above scheme on the homogenized network of a tree
network with heterogeneous links.  Similar to the unicast network, we
can show that this BATS code scheme with BP decoding can achieve a
multicast rate very close to the multicast capacity of a tree network
when $M$ is sufficiently large.

As another example, we apply Scheme~\ref{sch:6} on the three-layer
network with homogeneous links in Fig.~\ref{fig:threelayer}. We can
find three edge-disjoint trees in the network, each of which includes
one node in the middle layer and two destination nodes. The rank
distribution of each destination node can be characterized as in the
unicast network.  We can show (similar to the unicast network) that
Scheme~\ref{sch:6} with BP decoding can achieve a multicast rate very
close to the multicast capacity of the three-layer network when $M$ is
sufficiently large.

A three-layer network models a content distribution network (CDN),
where the source node is the server, the nodes in the middle layer are
the edge-servers that are close to the users represented by the nodes
in the bottom layer. The low storage requirement at the intermediate
nodes is a welcomed property for CDNs, where each edge-server in
parallel supports a lot of files.  BATS codes also have a good property
for caching in CDN networks. For example, each edge-server can cache a
number of batches of a file.  When a user wants to download the file,
it collects the batches of the file from one or more
edge-servers. Since a BATS code is an erasure correction code, caching
batches are more robust (subject to edge-server failure) and more
storage-efficient than caching replications of the original file.

\subsection{Bufferfly Network}

We use the butterfly network as an example to discuss how to design
better BATS code schemes for general multicast networks.  It is easy
to check that Scheme~\ref{sch:6} is not sufficient for the butterfly
network since only edge-disjoint trees are used.  Suppose that the
butterfly network is homogeneous.  We first propose a BATS code scheme
and then discuss some improvements of the scheme for a practical issue.

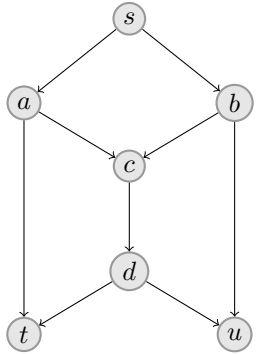
\begin{figure}
  \centering
  \begin{tikzpicture}[scale=1.4] 
	\node[dot] (s) at (0,0) {$s$}; 
	\node[dot] (a) at (-1,-0.8) {$a$} edge [<-] (s); 
	\node[dot] (b) at (1,-0.8) {$b$} edge [<-] (s); 
	\node[dot] (c) at (0,-1.4) {$c$} edge [<-] (b) edge [<-] (a); 
	\node[dot] (d) at (0,-2.4) {$d$} edge [<-] (c); 
	\node[dot] (t) at (-1,-3) {$t$} edge [<-] (d) edge [<-] (a); 
	\node[dot] (u) at (1,-3) {$u$} edge [<-] (d) edge [<-] (b); 
	\end{tikzpicture} 
  \caption{Butterfly network. Node $s$ is the source node. Node $t$
    and $u$ are destination nodes.}
  \label{fig:butterfly}
\end{figure}

\begin{scheme}[Butterfly network]
  \label{sch:butterfly1}
  In this scheme for the butterfly network, the source node $s$ generates
  batches of size $2M$ and transmits $M$ packets of a batch on link
  $(s,a)$ and the other $M$ packets of the batch on link
  $(s,b)$. Nodes $a$ and $b$ apply an inner code similar to that 
  defined in Scheme~\ref{sch:1} on batches of size $M$ except that
  they generate $2M$ coded packets for each batch so that they can
  transmit different packets on two of their outgoing links. Node $c$
  applies the inner code defined in Scheme~\ref{sch:1} with two
  exceptions: First node $c$ has a buffer size $2M-2$ packets since in
  each time slots it may receive two packets. Second, in each of the
  time slots $M, 2M, \ldots$, node $c$ generates $M$ coded packets
  using all the packets it has received in the current batch.  Node
  $d$ applies the inner code defined in Scheme~\ref{sch:1} for
  line networks on batches of size $M$ and transmits the same packets
  on both of their outgoing links.
\end{scheme}

Two places in the above scheme deserves some explanation. First,
node $a$ (or $b$) transmits different packets to its children, which
is different from the operation at node $d$. Note that both outgoing links of
node $a$ can reach node $t$. Therefore, transmitting different packets
on these two links can potentially increase the rank of the transfer
matrix of a batch. Second, node $c$ can potentially receive up to $2M$
packets of a batch, but it only generates $M$ recoded packets. This is
crucial to make the transmission on link $(c,d)$ efficient for both
destination nodes. %

In Scheme~\ref{sch:butterfly1}, the two destination nodes have the
same rank distribution for the transfer matrix of the batches.  We can
argue that for each destination node, the normalized expected rank of
the transfer matrix induced by this scheme converges to $1-\epsilon$
(using an approach similar to the one we have used for line networks and
unicast networks with homogeneous links), where $\epsilon$ is the link
erasure probability. Therefore, the BATS code scheme can achieve a
rate very close to the multicast capacity of the butterfly network
when $M$ is sufficiently large.

Now we incorporate a practical issue in the butterfly
network: consider that the two paths from node $s$ to node $c$ can
have different latencies.  With the latency issue, it is possible that
node $c$ receives packets from different batches on its incoming links
at the same time slot.  We propose two approaches to resolve the
latency issue in the butterfly network.

\begin{scheme}[Butterfly network]
  \label{sch:butterfly2}
  Consider the butterfly network with link latency.
  This scheme is the same as Scheme~\ref{sch:butterfly1} except that node
  $c$ has a larger buffer so that the packets of the same batch can be
  aligned in node $c$ for network coding.
\end{scheme}

The above scheme is feasible when the latency difference is in certain
measurable range.  For example, if we know that the two paths from
node $s$ to node $c$ have a maximum latency difference of $10$ time slots,
we can enlarge the buffer in node $c$ to cache $20$ more packets.  A
more robust scheme is still desired for the scenarios where the 
latency difference is large or not measurable.

\begin{scheme}[Butterfly network]
  \label{sch:butterfly4}
  Consider the butterfly network with link latency.  The source node
  separates its input packets into two groups $A$ and $B$. The source
  node encodes packets in $A$ and in $B$ using two BATS codes of batch
  size $M$. The batches belonging to group $A$ and $B$ are transmitted
  on link $(s,a)$ and $(s,b)$, respectively. Nodes $a$, $b$ and $d$
  apply the same inner code defined in Scheme~\ref{sch:butterfly1}.
  Node $c$ applies the same operation as node $a$ in
  Scheme~\ref{sch:4} for the two-way relay network.  Node $t$ first
  decodes the packets in group $A$ using the batches received from link
  $(a,t)$. The packets received from link $(d,t)$ are in batches of
  the form
  \begin{equation*}
    Y = X_A H_A + X_BH_B,
  \end{equation*}
  where $X_A$ and $X_B$ are batches belonging to groups $A$ and $B$,
  respectively, and $H_A$ and $H_B$ are the corresponding transfer matrices.  Since
  group $A$ has been decoded, node $t$ can recover the batch $X_A$ and
  cancel the effect of $X_A$ from the received batch $Y$. Then node
  $t$ decodes the packets in $B$.
  Node $u$ applies the same decoding procedure.
\end{scheme}

The butterfly network has two sub-trees with the node $s$ as the root
and nodes $t$ and $u$ as the leaves: one sub-tree includes nodes $a$,
$c$ and $d$; and the other sub-tree includes nodes $b$, $c$ and $d$.
In the above scheme, for each group of the input packets, we apply a
BATS code scheme for multicast in one of the two sub-trees. Since the
two sub-trees share the network link $(c,d)$, the batches of these two
BATS codes are mixed together to share the network link $(c,d)$.  Note
that we do not mix the batches of the same BATS code.  The decoding in
a destination node is a kind of successive cancellation: One group of
the input packets is first decoded using BP decoding of BATS codes;
The effect of this group is cancelled out from the mixed batches; The
other group of the input packets is then decoded using BP decoding of
BATS codes.

\subsection{Peer-to-peer Networks}

Consider a model of peer-to-peer (P2P) network with $N$ peers each of
which can send a packet to another peer randomly selected in a time
slot. A source node has a file to be distributed to the $N$ peers, and
can transmit one packet to a randomly selected peer in a time slot.
Both network coding and fountain codes have been considered for such
P2P networks \cite{Champel09,deb06}. BATS codes
share the same advantages of fountain codes in peer-to-peer networks,
and has the benefit of network coding in P2P networks.

In traditional chunk-based P2P networks, if a chunk becomes rare in
the network, the completion time of the file distribution will be
significantly extended.  When BATS codes are used in P2P networks, the
source node encodes the file using a BATS code and distributes the batches
in the P2P network.  Since all the batches are statistically identical,
there is no ``rare batch'' issue. A batch is useful for a peer,
as long as the batch has packets that are linearly independent of
those packets of the batch that have been received. The source node can
keep on injecting new batches into the network so that a peer can
easily obtain useful batches from other peers.

To decode the file in a BATS code-based P2P scheme, we require a peer to
receive a sufficient number of batches subject to certain rank
distribution constraints for the transfer matrices of the batches.
Instead of short completion time, peers are also hope to have low
receiving overhead.  When a peer has received more than one batch,
it performs \emph{batch selection} to determine which batch to
transmit. We can randomly select a batch but it may result in an issue
similar to the issue of random scheduling of chunks which we have
discussed in \ref{sec:previous}.  A better approach is the
\emph{priority selection} introduced by Sanghavi et
al. \cite{sanghavi07} for gossiping with multiple messages, where the
most recently received message is selected for sharing.  To share a
batch selected, instead of sharing the received packets of the batch,
a peer can transmit new packets coded from the received packets of the
batch.  The benefit of random linear coding in distributing a set of
 packets in a P2P network has been studied by Deb et al. \cite{deb06}.
In a decentralized and blind manner of transmission, simulations show
that random linear coding significantly reduces the receiving overhead
and achieves performance close to a centralized protocol.  Readers are
referred to \cite{he14} for the simulation and analysis of a BATS
code-based P2P scheme.

After a peer successfully decodes the file, the peer can generate and
transmit new batches to other peers as a seed. The batches generated
by the seed will be useful with high probability to all the other
peers which have not decoded the file. To
save storage space, a seed can delete the file and keep only a number
of randomly generated batches for sharing with other peers.

\section{Concluding Remarks}
\label{sec:conc}

Benefiting from previous research on network coding and fountain
codes, BATS codes are proposed as a rateless code for transmitting
files through multi-hop communication networks with packet
loss. In addition to low encoding/decoding complexity, BATS codes can be
realized with constant computation and storage complexity at the
intermediate nodes.  This desirable property makes BATS code a
suitable candidate for the making of universal network coding based
network devices that can potentially replace routers.

Our study in this paper provides the tools to optimize the performance
of BATS code and the guidelines to design BATS code-based network
communication schemes.  Examples of BATS code applications are given
for line networks and its extensions, general unicast networks and
several multicast networks. For general multicast networks, schemes
based on BATS codes can be developed as illustrated for the butterfly
network and the P2P networks. More work is expected to explore the
applicability of BATS codes and to study the implementation of BATS
code-based network communication systems.

\section*{Acknowledgements}

We thank Prof. R\"{u}diger Urbanke for his insightful input to this
work. We thank Cheuk Ting Li and Tom M. Y. Lam for their works in C/C++
implementation of BATS codes.

\appendices

\section{Incomplete Beta Function}
\label{sec:beta}

Beta function with integer parameters is used extensively in this
work. Related results are summarized here. For positive integer $a$ and $b$, the \emph{beta function} is defined by
\begin{align*}
  B(a,b) = \int_0^1t^{a-1}(1-t)^{b-1}\diff t = \frac{(a-1)!(b-1)!}{(a+b-1)!}.
\end{align*}
The \emph{(regularized) incomplete beta function} is defined as
\begin{align}
  \Bi_{a,b}(x) & = \frac{\int_0^xt^{a-1}(1-t)^{b-1}\diff t}{B(a,b)} \label{eq:incbetadd} \\
   & = \sum_{j=a}^{a+b-1} \binom{a+b-1}{j} x^j(1-x)^{a+b-1-j}. \nonumber
\end{align}
For more general discussion of beta functions, as well as incomplete
beta functions, please refer to \cite{handbook}.

Using the above definitions, we can easily show that
\begin{align}
  \int_0^1 \Bi_{a,b}(x) \diff x %
  & = \frac{b}{a+b}, \label{eq:dds}
\end{align}
and 
\begin{align}\label{eq:dsa}
    \Bi_{a+1,b}(x) = \Bi_{a,b}(x) - \frac{x^a(1-x)^b}{a B(a,b)}. 
\end{align}

\begin{lemma}\label{lemma:2d}
 $\frac{\Bi_{a+1,b}(x)}{\Bi_{a,b}(x)}$ is monotonically increasing in $x$.
\end{lemma}
\begin{IEEEproof}
 By \eqref{eq:dsa},
 \begin{IEEEeqnarray*}{rCl}
  \frac{\Bi_{a+1,b}(x)}{\Bi_{a,b}(x)}
  & = & 1 -\frac{x^{a}(1-x)^{b}}{aB(a,b) \Bi_{a,b}(x)} \\ 
  &	= & 1 - \frac{1}{aB(a,b) \sum_{j=a}^{a+b-1} \binom{a+b-1}{j} x^{j-a} (1-x)^{a-1-j}} \\ 
  & = & 1- \frac{1}{aB(a,b) \sum_{j=0}^{b-1} \binom{a+b-1}{j+a} x^{j} (1-x)^{-1-j}}, 
  \end{IEEEeqnarray*}
  in which $ x^{j} (1-x)^{-1-j}$ is monotonically increasing.
\end{IEEEproof}

\begin{lemma}\label{lemma:2c}
  When $\frac{b-1}{a+1} \leq \frac{\eta}{1-\eta}$ where
  $0<\eta<1$, $\frac{\Bi_{a+1,b}(x)}{\Bi_{a,b}(x)}\leq 1-
  \frac{\eta}{b}$ for $0<x\leq 1-\eta$ with equality when
  $b=1$ and $x=1-\eta$.
\end{lemma}
\begin{IEEEproof}
 Since $\frac{\Bi_{a+1,b}(x)}{\Bi_{a,b}(x)}$ is monotonically increasing in $x$ (cf. Lemma~\ref{lemma:2d}), 
  it is sufficient to show $\frac{\Bi_{a+1,b}(1-\eta)}{\Bi_{a,b}(1-\eta)}\leq 1- \frac{\eta}{b}$. 
  Since $a+1\geq (b-1)\frac{1-\eta}{\eta}$, 
  \begin{align*}
   \Bi_{a,b}(1-\eta) & = \sum_{j=a}^{a+b-1} \binom{a+b-1}{j} (1-\eta)^{j} \eta^{a+b-1-j} \\
    & \leq b \binom{a+b-1}{a} (1-\eta)^{a} \eta^{b-1},
  \end{align*}
  where the equality holds for $b=1$.
  Thus, 
 \begin{align*}
  \frac{\Bi_{a+1,b}(1-\eta)}{\Bi_{a,b}(1-\eta)} & = 1- \frac{(1-\eta)^{a}\eta^{b}}{aB(a,b) \Bi_{a,b}(1-\eta)} \\ &  
	\leq 1 - \frac{(1-\eta)^{a}\eta^{b}}{abB(a,b) \binom{a+b-1}{a} (1-\eta)^{a} \eta^{b-1}}  \\ & 
	= 1 - \frac{\eta}{b}.
  \end{align*}
\end{IEEEproof}

We will use the following result about the summation of binomial coefficients: 
\begin{equation}\label{lemma:kda}
  \sum_{j=0}^n (-1)^{j-n}\binom{j+m}{n}\binom{n}{j} = 1, \quad m\geq n.
\end{equation}
The above equality can be verified as follows:
  \begin{IEEEeqnarray*}{rCl}
    \sum_{j=0}^n (-1)^{j-n}\binom{j+m}{n}\binom{n}{j}
    & = & \sum_{j=0}^n (-1)^{j-n}\binom{j+m}{j+m - n}\binom{n}{j} \\ 
    & = & \sum_{j=0}^n (-1)^{j-n}(-1)^{j+m-n} %
    \binom{-j-m+j+m-n-1}{j+m - n}\binom{n}{j} \IEEEyesnumber \label{eq:neg1} \\ 
    & = & \sum_{j=0}^n (-1)^{m}\binom{-n-1}{j+m - n}\binom{n}{n-j} \\ 
    & = & (-1)^{m} \binom{-1}{m}\IEEEyesnumber \label{eq:vand} \\ 
    & = & 1,\IEEEyesnumber \label{eq:neg2}
  \end{IEEEeqnarray*}
  where \eqref{eq:vand} follows from Vandermonde's identity; 
  \eqref{eq:neg1} and \eqref{eq:neg2} use the relation between
  binomial coefficients with negative integers and positive integers.

\begin{lemma}\label{lemma:cs}
For $r\geq 1$,
  $$\sum_{d=r+1}^{\infty}\frac{1}{d-1}\Bi_{d-r,r}(x) = -\ln(1-x), \quad x\in [0,\ 1). $$
\end{lemma}
\begin{IEEEproof}
  As a special case, when $r=1$, the equality becomes
  \begin{equation}\label{eq:cs-1}
    \sum_{d=2}^{\infty} \frac{x^{d-1}}{d-1} = - \ln(1-x),
  \end{equation}
  which is the Taylor expansion of $-\ln(1-x)$ for $x\in [0, 1)$.

  To prove the general case, let us first derive an alternative form of $\Bi_{d-r,r}(x)$. For $a>0$,
  \begin{IEEEeqnarray*}{rCl}
    \Bi_{a,b}(x)
   & = & \sum_{j=a}^{a+b-1} \binom{a+b-1}{j}x^j \sum_{i=0}^{a+b-1-j}(-1)^i\binom{a+b-1-j}{i}x^i \\ 
   & = & \sum_{m=a}^{a+b-1}x^m \sum_{j=a}^m \binom{a+b-1}{j} (-1)^{m-j} \binom{a+b-1-j}{m-j} \\ 
   & = & \sum_{m=a}^{a+b-1}(-x)^m \binom{a+b-1}{m} \sum_{j=a}^m \binom{m}{j} (-1)^{j}  \\ 
   & = & \sum_{m=a}^{a+b-1}(-x)^m \binom{a+b-1}{m} \binom{m-1}{a-1} (-1)^a \\ 
   & = & b \binom{a+b-1}{b}(-1)^a \sum_{m=a}^{a+b-1}\frac{(-x)^m}{m} \binom{b-1}{m-a}.
   \end{IEEEeqnarray*}
   Using this form for $\Bi_{d-r,r}(x)$, we have
   \begin{IEEEeqnarray*}{rCl}
     \sum_{d=r+1}^{\infty}\frac{1}{d-1}\Bi_{d-r,r}(x)
     & = & \sum_{d=r+1}^{\infty}\frac{r}{d-1} \binom{d-1}{r}\sum_{m = d-r}^{d-1} \binom{r-1}{m-d+r} (-1)^{m-d+r} \frac{x^m}{m} \\  
     & = & \sum_{m=1}^{\infty} \frac{x^m}{m}  \sum_{d=\max\{m,r\}+1}^{m+r}\frac{r}{d-1} \binom{d-1}{r} \binom{r-1}{m-d+r}(-1)^{m-d+r} \\ 
     & = & \sum_{m=1}^{\infty}\frac{x^m}{m} A_m, \IEEEyesnumber \label{eq:cs-2}
   \end{IEEEeqnarray*}
   where 
   \begin{equation*}
     A_m \triangleq \sum_{d=\max\{m,r\}+1}^{m+r}\frac{r}{d-1} \binom{d-1}{r} \binom{r-1}{m-d+r}(-1)^{m-d+r}.
   \end{equation*}
   For $m\leq r$,
   \begin{align*}
     A_m & 
     = \sum_{d=r+1}^{m+r} \frac{r}{d-1}\binom{d-1}{r} \binom{r-1}{m-d+r}(-1)^{m-d+r} \\ & 
     = \sum_{d=r+1}^{m+r} \binom{d-2}{r-1} \binom{r-1}{m-d+r}(-1)^{m-d+r} \\ &
     = \sum_{j=0}^{m-1} \binom{j+r-1}{r-1} \binom{r-1}{m-j-1}(-1)^{m-j-1} \\ &
     = \sum_{j=0}^{m-1} \binom{j+r-1}{m-1} \binom{m-1}{m-j-1}(-1)^{m-j-1} \\ &
     = 1, 
   \end{align*}
   where the last equality follows from~\eqref{lemma:kda}.
   Similarly, for $m>r$,
   \begin{align*}
     A_m & 
     = \sum_{d=m+1}^{m+r} \frac{r}{d-1}\binom{d-1}{r} \binom{r-1}{m-d+r}(-1)^{m-d+r} \\ &
     = \sum_{d=m+1}^{m+r} \binom{d-2}{r-1} \binom{r-1}{m-d+r}(-1)^{m-d+r} \\ &
     = \sum_{j=0}^{r-1} \binom{j+m-1}{r-1} \binom{r-1}{r-j-1}(-1)^{r-j-1} \\ &
     = 1.
   \end{align*}
   The proof is completed by referring to \eqref{eq:cs-1} and \eqref{eq:cs-2} with $A_m=1$.
\end{IEEEproof}

\section{Layered Decoding Graph}
\label{sec:layered}

We have discussed different decoding strategies under the rule that a
check node is decodable if and only if its rank equals its degree.
We say a variable node is decodable if it is connected to a decodable
check node.  In Section~\ref{sec:dec}, a decodable check node is
chosen and all its neighbors (variable nodes) are recovered
simultaneously, while in Section~\ref{sec:rd}, a decodable variable node is
uniformly chosen to be recovered.  Here we show that under the decoding
rule that a check node is decodable if and only if its rank equals
its degree, both strategies stop with the same subset of the variable
nodes undecoded.

For a given decoding graph $\mathcal{G}$, let $\mathcal{G}^{0} =
\mathcal{G}$. Label by $L_1$ all the decodable check nodes in
$\mathcal{G}^{0}$ and label by $L_2$ all the variable nodes in
$\mathcal{G}^{0}$ connected to the check nodes with label $L_1$. We
repeat the above procedure as follows. For $i=1,2,\ldots$, let
$\mathcal{G}^{i}$ be the subgraph of $\mathcal{G}$ obtained by
removing all the nodes with labels $L_j$ for $j\leq 2i$, as well as
the adjacent edges. (The generator matrices of the check nodes are also
updated.) Label by $L_{2i+1}$ all the decodable check nodes in
$\mathcal{G}^{i}$ and label by $L_{2i+2}$ all the variable nodes in
$\mathcal{G}^{i}$ connected to the check nodes with label
$L_{2i+1}$. This procedure stops when $\mathcal{G}^{i}$ has no more
decodable check nodes. Let $i_0$ be the index where the procedure
stops. The above labelling procedure is deterministic and generates
unique labels for each decodable variable nodes and check nodes.

With the labels, we can generate a layered subgraph $\mathcal{G}'$ of
$\mathcal{G}$.  In $\mathcal{G}'$, layer $j$, $j=1,2,\ldots,2i_0$,
contains all the check/variable nodes with label $L_j$. Only the edges
connecting two nodes belonging to two consecutive layers are preserved in
$\mathcal{G}'$. By the assigning rule of the labels, it is clear that
a variable node on layer $2i$ must connect to one check node on layer
$2i-1$, $i=1,\ldots,i_0$, because otherwise the variable node is not
decodable. Further, a check node on layer $2i+1$ must connect to some
variable nodes on layer $2i$, $i=1,\ldots,i_0-1$, because otherwise the
check node should be on layer $2i-1$.

By the definition of decodability, a decoding strategy must process
the variable/check nodes in $\mathcal{G}'$ following an order such that a
variable/check node is processed after all its lower layer descendant
variable/check nodes have been processed.  The two random decoding
strategies we have discussed in Section~\ref{sec:dec} and
Section~\ref{sec:rd} both can process all the nodes in $\mathcal{G}'$ before
stopping.

\section{Solving the system of differential equations}
\label{sec:sol}

We solve the following system of differential equations given in
\eqref{eq:df1} and \eqref{eq:df2}, which is reproduced as follows:
\begin{IEEEeqnarray*}{rCl}
  \frac{d\rho_{d,r}(\tau)}{d \tau}
  & = & (\alpha_{d+1,r}\rho_{d+1,r}(\tau)+\bar
  \alpha_{d+1,r+1}\rho_{d+1,r+1}(\tau)-\rho_{d,r}(\tau)) \frac{d}{\theta -\tau},  \\
  & & \quad 1\leq r \leq M,\ r< d\leq D,  \\
  \frac{d \rho_0(\tau)}{d \tau} &  = & \frac{\sum_{r=1}^{D-1}r\alpha_{r+1,r}
    \rho_{r+1,r}(\tau) -  \rho_0(\tau)}{\theta -\tau} - 1
\end{IEEEeqnarray*}
with initial values $\rho_{d,r}(0)$ and $ \rho_0(0)$.

Let $y_{d,r}(\tau)=(1-\tau/\theta)^{-d} \rho_{d,r}(\tau)$. We have %
\begin{equation*}
  \frac{d y_{d,r}(\tau)}{d \tau}  = \frac{d}{\theta}(\alpha_{d+1,r}y_{d+1,r}(\tau)+\bar \alpha_{d+1,r+1}y_{d+1,r+1}(\tau)).
\end{equation*}
We see that $y_{d,r}(0) = \rho_{d,r}(0)$.
Define
\begin{align*}
  \hat \rho_{d,r}^{(0)} & = \rho_{d,r}(0) \\
  \hat \rho_{d,r}^{(i+1)} & = \alpha_{d-i,r}\hat \rho_{d,r}^{(i)} + \bar \alpha_{d-i,r+1} \hat \rho_{d,r+1}^{(i)}.
\end{align*}
We can verify that %
\begin{align*}
  y_{d,r}(\tau) = \sum_{j = d}^D \binom{j-1}{d-1} (\tau/\theta)^{j-d}\hat \rho_{j,r}^{(j-d)}.
\end{align*}
Thus %
\begin{align}
  \rho_{d,r}(\tau) = (1-\tau/\theta)^{d}\sum_{j = d}^D \binom{j-1}{d-1} (\tau/\theta)^{j-d}\hat \rho_{j,r}^{(j-d)}. \label{eq:so8d}
\end{align}

Using the general solution of linear differential equations, we obtain
that
\begin{IEEEeqnarray*}{rCl}
  \rho_0(\tau)  
  & = & (1-\tau/\theta) %
  \Bigg(\int_{0}^{\tau}\frac{\sum_{r = 1}^M r \alpha_{r+1,r}\rho_{r+1,r}(t)}{\theta-t}(1-t/\theta)^{-1} dt %
  + \theta \ln(1-\tau/\theta) +  \rho_{0}(0) \Bigg) \\ 
& = & 
(1-\tau/\theta) %
\Bigg(\sum_{r = 1}^M r\alpha_{r+1,r}\int_{0}^{\tau}\frac{ \rho_{r+1,r}(t)}{\theta-t}(1-t/\theta)^{-1} dt %
+ \theta \ln(1-\tau/\theta) + \rho_{0}(0) \Bigg). \IEEEyesnumber \label{eq:dff-1}
\end{IEEEeqnarray*}
The integral in \eqref{eq:dff-1} can be further calculated as follows: 
\begin{IEEEeqnarray*}{rCl}
  \IEEEeqnarraymulticol{3}{l}{\int_{0}^{\tau}\frac{\rho_{r+1,r}(t)}{\theta-t}(1-t/\theta)^{-1} dt} \\ \   
  & = & \label{eq:diff-2}
  \int_{0}^{\tau}\frac{\sum_{j = r+1}^D\hat \rho_{j,r}^{(j-r-1)} \binom{j-1}{r} (1-t/\theta)^{r+1}(t/\theta)^{j-r-1}}{(\theta-t)(1-t/\theta)} dt \\ 
  & = & 
  \int_{0}^{\tau}{\sum_{j = r+1}^D \hat \rho_{j,r}^{(j-r-1)} \binom{j-1}{r} (1-t/\theta)^{r-1}(t/\theta)^{j-r-1}} \frac{dt}{\theta} \\ 
  & = & 
  \sum_{j = r+1}^D\hat \rho_{j,r}^{(j-r-1)} \binom{j-1}{r} \int_{0}^{\tau/\theta} (1-t)^{r-1}t^{j-r-1} dt \\
   & = & \IEEEyesnumber \label{eq:diff-3}
   \sum_{j = r+1}^D \hat \rho_{j,r}^{(j-r-1)} \binom{j-1}{r} \frac{(j-r-1)!(r-1)!}{(j-1)!} \Bi_{j-r,r}(\tau/\theta) \\ 
  & = & 
  1/r \sum_{j = r+1}^D \hat \rho_{j,r}^{(j-r-1)}  \Bi_{j-r,r}(\tau/\theta),
\end{IEEEeqnarray*}
where \eqref{eq:diff-2} is obtained by substituting $\rho_{r+1,r}(t)$
in \eqref{eq:so8d}, and \eqref{eq:diff-3} is obtained by the
definition of incomplete beta function (cf. \eqref{eq:incbetadd}).

\end{document}